\documentclass[aps,pra,twocolumn,amsmath,amssymb,showpacs,floatfix,superscriptaddress]{revtex4-1}

\usepackage{times}
\usepackage[utf8x]{inputenc}
\usepackage[english]{babel}
\usepackage[T1]{fontenc}
\usepackage{lmodern}
\usepackage{amssymb}
\usepackage{amsmath}
\usepackage{amsthm}
\usepackage[pdftex]{graphicx}
\usepackage{epstopdf}
\usepackage{braket}
\usepackage[bottom]{footmisc}
\usepackage{dcolumn}
\usepackage{bm}
\usepackage{color}
\usepackage{relsize,dsfont,mathrsfs,empheq,verbatim,upgreek}
\usepackage{etoolbox}
\usepackage[caption = false]{subfig}

\usepackage{enumerate}

\newcommand{\Tr}{\mathrm{Tr}}
\newcommand{\Lt}{\mathcal{L}}

\begin{document}


\title{Dynamically Emergent Quantum Thermodynamics: 
Non-Markovian Otto Cycle}


\author{Irene Ada Picatoste}

\affiliation{Institute of Physics, University of Freiburg, 
Hermann-Herder-Stra{\ss}e 3, D-79104 Freiburg, Germany}

\author{Alessandra Colla}

\affiliation{Institute of Physics, University of Freiburg, 
Hermann-Herder-Stra{\ss}e 3, D-79104 Freiburg, Germany}

\author{Heinz-Peter Breuer}

\affiliation{Institute of Physics, University of Freiburg, 
Hermann-Herder-Stra{\ss}e 3, D-79104 Freiburg, Germany}

\affiliation{EUCOR Centre for Quantum Science and Quantum Computing,
University of Freiburg, Hermann-Herder-Stra{\ss}e 3, D-79104 Freiburg, Germany}

\begin{abstract}
Employing a recently developed approach to dynamically emergent quantum thermodynamics, we revisit the 
thermodynamic behavior of the quantum Otto cycle with a focus on memory effects and strong system-bath 
couplings. Our investigation is based on an exact treatment of non-Markovianity by means of an exact
quantum master equation, modelling the dynamics through the Fano-Anderson model featuring
a peaked environmental spectral density. By comparing the results to the standard Markovian case, we find that 
non-Markovian baths can induce work transfer to the system, and identify specific parameter regions which lead to 
enhanced work output and efficiency of the cycle. In particular, we demonstrate that these improvements 
arise when the cycle operates in a frequency interval which contains the peak of the spectral density.
This can be understood from an analysis of the renormalized frequencies emerging through the system-baths couplings.
\end{abstract}

\date{\today}

\maketitle

\section{Introduction}\label{sec:Intro}

Quantum thermodynamics is currently a prolific and exciting research field. Technical advancements, the precision of 
controlling quantum systems in the lab, and the chase for the quantum advantage, have led to the fast-paced development 
of quantum thermal machines, insofar as to already bring experimental realization 
\cite{Abah2012,Rossnagel2016,deAssis2019,Lindenfels2019,Peterson2019,Myers2022}.
Yet, because of this accelerated growth, the field is also plagued with inconsistencies: quantum engines and other 
applications are being theoretized, but essential foundations of the field's theory are still under debate.

While the formulation and fundamentals of quantum thermodynamics seem to be well understood for ultra-weak system-bath coupling and continuous information loss, the community is still not converging on basic definitions for thermodynamic quantities outside of this regime.  There are many aspects that make the quest challenging, but we can perhaps identify two major contributors. First, the interaction energy between a quantum system and its environment becomes non-negligible at stronger coupling parameters. It therefore needs to be understood how much of this energy is accessible from the system and should be counted as part of the system internal energy, and whether (or how) it counts as work or heat. The second aspect is the complex information exchange between system and environment. It is now clear that information theory and thermodynamics share a deeper link and that irreversible behaviour is connected to information loss; thus, memory effects and information backflow from the bath to the system should also significantly influence the theory \cite{Breuer2016a}. 

In the literature one may find numerous theoretical proposals addressing these questions \cite{Weimer2008,Esposito2010,Teifel2011,Alipour2016,Seifert2016,Strasberg2017,Rivas2020,Alipour2021,Landi2021}. However, few of these disparate approaches made it so far as to enter the stage of theoretical applications. Perhaps the most popular strong coupling treatment of quantum thermodynamics is given by identifying the heat exchanged as minus the energy change of the bath, as proposed by Esposito, Lindenberg and Van den Broeck (ELB) \cite{Esposito2010}. This corresponds to assigning the interaction energy fully to the system's internal energy, in the original formulation. Later treatments have instead kept the definition of internal energy as in the weak coupling formulation (i.e.  excluding the interaction energy) while adopting the ELB definition for heat, imposing some work exchange \textit{ad hoc} to maintain a first law of thermodynamics \cite{Landi2021}. This definition of heat exchange seems therefore too arbitrary a choice; moreover, it does not seem to reflect the impact of memory effects, which could arise from finite heat baths or structured spectral densities. The oscillations of entropy production according to this definition, i.e.  temporary violations of the second law, seem in fact to have no relation with non-Markovian behavior \cite{Strasberg2019}. Further, this definition of heat exchange has been found to be incompatible with the weak coupling formulation in the limit of vanishing coupling parameters when the system is driven by an external time-dependent force \cite{Colla2021}. 

Already a vast quantity of studies emerged in very recent years, that are concerned with investigating strong coupling and/or non-Markovian effects on a quantum Otto cycle \cite{Zhang2014,Pozas-Kerstjens2018,Thomas2018,Pezzutto2019,Mukherjee2020,Wiedmann2020,Wiedmann2021,Chakraborty2022,Kaneyasu2023,Ishizaki2023,Liu2021}.
While these works explore the impact of different causes of non-Markovianity and find an array of interesting consequences, most of them either employ the weak-coupling treatment of quantum thermodynamics or the ELB approach within an exact treatment.  Because of the reasons explained above, we believe some features might have been missed. We hope to help fill this gap by properly accounting for strong coupling and non-Markovian effects,  starting by employing more suitable definitions of the main thermodynamic quantities.

We have recently proposed a different approach to the thermodynamic laws that is tailored to exactly resolve the consequences of coupling a system to an arbitrary reservoir \cite{Colla2022a}. This arbitrariness can refer to the coupling strength, to the protocol for switching on and off the interaction,  the finiteness of the bath or to its structural properties, including its initial state. This approach particularly was shown to link violations of the second law with information backflow, and thus represents in our opinion the best formulation in which to study the properties of a non-Markovian thermal machine.  In Ref.~\cite{Colla2022a} the thermodynamic quantities are given in terms of an effective system Hamiltonian $K_S$ emerging from the dynamics of the system when it is coupled with the environment. This emergent Hamiltonian contains details of the environment and total evolution, and describes how the system energy levels react to the interaction with the environment. Accordingly, the internal energy of the system can change either due to a change in the energy levels or eigenstates of the effective Hamiltonian (work), or due to a change of the populations (heat).
It turns out that non-Markovian environments can in general induce a time-dependent effective Hamiltonian $K_S(t)$ 
even when the total system (system+environment) is an isolated system (following a unitary evolution according to a 
time-independent Hamiltonian) and even for weak coupling parameters. This reveals a major characteristic of 
non-Markovian baths: They can perform work on the system, which is interpreted as useful, coherent energy granted to 
(or taken away from) the system as a result of non-trivial information exchange with the environment.  This can of course lead to crucial changes in the treatment of heat engines and is the main motivation behind this work.

The focus of this paper is the variation in work output and efficiency of a heat engine when the baths can be strongly coupled and exhibit memory effects. We do this by comparing the standard Otto cycle -- a harmonic oscillator driven and coupled to Markovian baths \cite{Kosloff2017} -- with the corresponding cycle where the system-baths couplings are
taken into account within an exact non-Markovian treatment. To this end, the system-baths dynamics are described
by the Fano-Anderson model \cite{Fano1961,Anderson1961,Mahan2000}
featuring a structured spectral density. We note that in 
Refs.~\cite{Huang2022a,Huang2022b} an approach to quantum thermodynamics of the Fano-Anderson model has been developed which is closely related to our approach.

The paper is organized as follows.
In Sec.~\ref{sec:theory} we collect the theoretical background needed for this study: We summarize in 
Sec.~\ref{sec:qthermo} the approach for quantum thermodynamics developed in \cite{Colla2022a} and present some
details of the microscopic model and its solution in Sec.~\ref{sec:micro-model}.  In Sec.~\ref{sec:otto-cycle}, we show the theoretical results for the Otto cycle where the system is connected via arbitrary coupling parameters to two thermal baths of harmonic oscillators with peaked Lorentzian spectral density.
Since the coupling with these baths leads in general to both heat and work exchange, we factor this in for the engine's work output and efficiency. In Sec.~\ref{sec:numerics} we show numerical results in different parameter regimes and find that if the bath spectral density peaks at a frequency within the driving range of the system, then the cycle can be enhanced, showing higher efficiencies than its standard counterpart. Finally, we draw our conclusions in Sec.~\ref{Conclu}.

\section{Theory}\label{sec:theory}

\subsection{Nonequilibrium thermodynamics at strong couplings}\label{sec:qthermo}

We consider the nonequilibrium quantum thermodynamics of an open system $S$ coupled 
to an environment $E$ representing the heat baths. The quantum states of $S$ and $E$ are
given by density matrices $\rho_S$ and $\rho_E$, respectively, while states of the total
composite system $S+E$ are denoted by $\rho_{SE}$. The microscopic Hamiltonian is taken 
to be of the form
\begin{equation} \label{ham-total}
 H(t) = H_S(t) + H_E + H_I(t),
\end{equation}
where $H_S$ and $H_E$ represent the system and the environmental Hamiltonian, respectively, and
$H_I$ denotes the interaction Hamiltonian. Note that the system Hamiltonian and the interaction Hamiltonian
may be time dependent, considering, e.g., an external driving of the system or a switching 
on and off of the system-environment interaction. 

Our approach to nonequilibrium thermodynamics at arbitrary system-environment coupling
developed in \cite{Colla2022a} is based on the following master equation for 
the density matrix $\rho_S(t)$ of the open system $S$:
\begin{equation} \label{tcl-meq}
 \frac{d}{dt}\rho_S(t) = \Lt_t[\rho_S(t)] = -i \left[K_S(t),\rho_S(t)\right] + {\mathcal{D}}_t  [\rho_S(t)].
\end{equation}
This is an exact time-local differential equation with an explicitly time-dependent 
generator $ \Lt_t$ which completely describes all memory effects of the open system dynamics 
\cite{Shibata1977,Shibata1979}. Since there is no 
time-convolution in this master equation, it is also known as {\it{time-convolutionless master equation}}.
The generator of the master equation \eqref{tcl-meq} consists of two parts. 
The first one is a Hamiltonian contribution given by the commutator with some effective Hamiltonian
$K_S(t)$, while the second contribution is provided by the dissipator
\begin{equation} \label{dissipator}
 {\mathcal{D}}_t [\rho_S] = \sum_{k}\gamma_{k}(t)\Big[L_{k}(t)
  \rho_S L_{k}^{\dag}(t) - \frac{1}{2}\big\{L_{k}^{\dag}(t)L_{k}(t),\rho_S\big\}\Big]
\end{equation}
with certain decay rates $\gamma_k(t)$ and corresponding Lindblad operators $L_k(t)$, which are in general time-dependent. The above general structure of the time-local master equation can be
derived from the requirements of the preservation of Hermiticity and trace of the density matrix 
\cite{Andersson2014a,Breuer2012a}. 

The explicit form 
of the rate functions $\gamma_k(t)$ and of the Lindblad operators $L_k(t)$ for a given microscopic
Hamiltonian can be found by means of a perturbation expansion in the system-environment coupling by applying 
the time-convolutionless projection operator technique \cite{Shibata1977,Shibata1979,Breuer2002}. 
This expansion can also be viewed as an expansion of the generator in terms of so-called ordered cumulants
\cite{Kampen1974a,Kampen1974b,Breuer2002}.
In this paper, we will investigate an analytically solvable system-reservoir 
model for which the exact master equation \eqref{tcl-meq} is known \cite{Tu2008,Jin2010,Lei2012,Zhang2012}
and can be constructed from the solution of the full Heisenberg equations of motion (see Sec.~\ref{sec:micro-model}).

Given the master equation \eqref{tcl-meq} it seems natural to interpret the Hermitian operator appearing in the
commutator part as an effective system Hamiltonian. However, this interpretation has to be taken with care,
because the splitting of the generator $ \Lt_t$ into Hamiltonian and dissipative part is not unique. In other words,
one can modify the commutator part and, at the same time, the dissipator part in \eqref{tcl-meq} in an
infinite number of different ways without changing the generator itself and, hence, without changing the 
dynamics of the system. To fix a unique splitting of the generator into Hamiltonian contribution and dissipator
we have proposed a {\it{principle of minimal dissipation}} \cite{Colla2022a} which is based on the approach 
developed in \cite{Sorce2022}.
This principle requires that the dissipator of the generator must be minimal with respect to a certain
superoperator norm. Physically this means that the dissipative contribution of the master equation corresponds 
to that part of the total energy flow which cannot be interpreted as mechanical work due to a time-dependent
effective Hamiltonian of the system and, thus, must describe the flow of heat energy. As shown in \cite{Sorce2022} 
the dissipator is minimal if and only if the Lindblad generators $L_k(t)$ are traceless, a condition which leads 
to a unique expression for the effective system Hamiltonian $K_S(t)$.

In accordance with these considerations we define the internal energy of the system by the expectation value of the effective Hamiltonian,
\begin{equation} \label{internal-energy}
 U_S (t) = \Tr \{K_S(t)\rho_S(t)\},
\end{equation}
and formulate the first law on the change of internal energy as
\begin{eqnarray}\label{first-law}
 \Delta U_S (t) \equiv U_S(t) - U_S(0) = W_S(t) + Q_S(t),
\end{eqnarray}
where we have defined work and heat contributions by
\begin{eqnarray}
 W_S(t) &=& \int_0^t d\tau \, \Tr \big\{ \dot{K}_S(\tau) \rho_S(\tau) \big\},  \label{work} \\
 Q_S(t) &=& \int_0^t d\tau \, \Tr \big\{ K_S(\tau) \dot{\rho}_S(\tau)  \big\}.  \label{heat}
\end{eqnarray}
Note that according to the master equation \eqref{tcl-meq} the heat contribution $Q_S(t)$ can be expressed
in terms of the dissipator of the master equation as
\begin{equation}
 Q_S(t) = \int_0^t d\tau \, \Tr \big\{ K_S(\tau) \mathcal{D}_{\tau}[\rho_S(\tau)] \big\}.
\end{equation}

\subsection{Microscopic model and effective Hamiltonian}\label{sec:micro-model}

Our goal is a study of the thermodynamics of the Otto cycle in which the full non-Markovian behavior
of the system-baths coupling is taken into account, including all memory effects due to strong couplings
and structured spectral densities.
To keep the analysis as simple as possible we consider an analytically solvable model known as
Fano-Anderson model 
\cite{Fano1961,Anderson1961,Mahan2000}, 
which can be described by the microscopic Hamiltonian
\begin{eqnarray} \label{total-ham}
 H &=& H_S + H_E + H_I \nonumber \\
 &=& \omega_0 a^{\dagger}a + \sum_j \omega_j  c_j^{\dagger}c_j 
 + \sum_j \left( g_j a^{\dagger}c_j + g^*_j ac_j^{\dagger} \right),
\end{eqnarray}
where $a^{\dagger}$, $a$ are the creation and annihilation operators of the open system mode with frequency 
$\omega_0$, and $c_j^{\dagger}$, $c_j$ are the creation and annihilation operators of the environmental modes
with frequencies $\omega_j$, satisfying the commutation relations
$[c_j,c^{\dagger}_k]=\delta_{jk}$.
The interaction Hamiltonian $H_I$ describes the coupling between the system mode
$a$ and the environmental modes $c_j$ with corresponding coupling constants $g_j$. As the total Hamiltonian
is quadratic, the model can easily be solved analytically to obtain the reduced dynamics of the system mode $a$. To this end, we define
the memory kernel
\begin{equation} \label{memory-kernel}
 \mathcal{K}(t,t_1) = \mathcal{K}(t-t_1) 
 = \sum_j |g_j|^2 e^{-i\omega_j (t-t_1)},
\end{equation}
which can also be written as
\begin{equation} \label{memory-kernel-2}
 \mathcal{K}(t,t_1) = \int_0^{\infty} d\omega J(\omega) 
 e^{-i\omega (t-t_1)},
\end{equation}
where we have introduced the spectral density
\begin{equation} \label{spectral-density}
 J(\omega) = \sum_j |g_j|^2 \delta(\omega-\omega_j),
\end{equation}
which will later be replaced by a continuous function.
Defining further the Green function $G(t)$ as the solution of the 
integro-differential equation
\begin{equation} \label{green-function}
 \frac{d}{dt}G(t) + i\omega_0 G(t) +  \int_0^t dt_1 \mathcal{K}(t-t_1) 
 G(t_1) = 0
\end{equation}
corresponding to the initial value $G(0)=1$, we obtain the following exact 
solution of the Heisenberg equations of
motion for the system mode,
\begin{equation} \label{Heisenberg-solution}
 a(t) = G(t) a(0) + \int_0^t dt_1 G(t-t_1) c(t_1).
\end{equation}
Here, $a(0)=a$ represents the Schr\"odinger picture operator and we have defined
\begin{equation} \label{def-c}
 c(t) = -i \sum_j g_j e^{-i\omega_j t} c_j(0),
\end{equation}
which involves the mode operators $c_j(0)=c_j$ of the environment in the Schr\"odinger picture.

Employing Eq.~\eqref{Heisenberg-solution} we can determine all relevant moments of the system mode $a$ and also an
exact time-local master equation for the open system's density matrix $\rho_S(t)$. For simplicity we assume
a factorizing total initial state \footnote{We remark that an extension
of our approach to correlated initial states seems possible by means of
the formalism developed in \cite{Colla2022b}}
\begin{equation} \label{rho-init}
 \rho_{SE}(0) = \rho_S(0) \otimes \rho_E(0),
\end{equation}
where the initial environmental state represents a Gibbs state
\begin{equation} \label{rho-gibbs}
 \rho_E(0) = \frac{1}{Z_E} e^{-\beta H_E}
\end{equation}
with $Z_E$ the partition function and $\beta=1/k_BT$ the inverse temperature. It follows from 
Eqs.~\eqref{rho-init} and \eqref{def-c} that the first and second moments of $c(t)$ vanish,
\begin{equation} \label{moments-c}
 \langle c(t) \rangle = 0, \qquad \langle c(t_1) c(t_2) \rangle = 0,
\end{equation}
and, moreover, that the cross correlations between $a(0)$ and $c(t)$ are zero,
\begin{equation} \label{cross-corr}
 \langle a(0) c(t) \rangle = 0, \qquad \langle a(0) c^{\dagger}(t) \rangle = 0.
\end{equation}
With the help of \eqref{Heisenberg-solution} we then find the following expressions for the first and second
moment of the open system mode,
\begin{eqnarray} 
 \langle a \rangle_t &=& G(t) \langle a \rangle_0, \label{moment-a} \\
 \langle aa \rangle_t &=& G^2(t) \langle aa \rangle_0, \label{moment-aa} \\
 \langle a^{\dagger}a \rangle_t &=& |G(t)|^2 \langle a^{\dagger}a \rangle_0 + I(t). \label{moment-a*a}
\end{eqnarray}
Here, we use the notation $\langle X \rangle_t = \Tr \{ X \rho_S(t)\}$ for any system operator $X$ and define the 
noise integral $I(t)$ through
\begin{equation} \label{noise-integral}
 I(t) = \int_0^t dt_1  \int_0^t dt_2 G^*(t-t_1) G(t-t_2) \langle c^{\dagger}(t_1) c(t_2) \rangle.
\end{equation}

From the expressions \eqref{moment-a}-\eqref{moment-a*a} for the moments of mode 
$a$ one can derive an exact time-convolutionless master equation for the open 
system density matrix \cite{Tu2008,Jin2010,Lei2012,Zhang2012}
which is of the form given in Eq.~\eqref{tcl-meq}
(see Appendix \ref{app-A} for details), where 
the effective Hamiltonian takes the form
\begin{equation} \label{K_S-fano-anderson}
 K_S(t) = \omega_r(t) a^{\dagger}a,
\end{equation}
and the dissipator reads
\begin{eqnarray} \label{dissipator-fano-anderson}
  {\mathcal{D}}_t[\rho_S] &=&  
 \gamma(t)(N(t)+1) \left[a\rho_Sa^{\dagger}-\frac{1}{2}\left\{a^{\dagger}a,\rho_S\right\}\right] \nonumber \\
 && + \gamma(t) N(t) \left[a^{\dagger}\rho_Sa-\frac{1}{2}\left\{aa^{\dagger},\rho_S\right\}\right].
\end{eqnarray}
Thus we see that the effective Hamiltonian  \eqref{K_S-fano-anderson} 
represents a harmonic oscillator with a renormalized frequency which is given by
\begin{equation} \label{def-omega_r}
 \omega_r(t) = \omega_0 + \Delta\omega(t) = - \Im \left( \frac{\dot{G}(t)}{G(t)} \right).
\end{equation}
The quantity $\gamma(t)$ plays the role of a generalized decay rate defined by
\begin{equation} \label{def-gamma}
 \gamma(t) = - 2 \Re \left( \frac{\dot{G}(t)}{G(t)} \right),
\end{equation}
and, finally, the quantity $N(t)$ is obtained from
\begin{equation} \label{def-N-t}
 N(t) = I(t) + \frac{\dot{I}(t)}{\gamma(t)}.
\end{equation}
Note that we have written the effective Hamiltonian \eqref{K_S-fano-anderson} 
and the dissipator \eqref{dissipator-fano-anderson} in a form which clearly reveals
the Born-Markovian approximation of the exact master equation. In fact, 
within second order in the system-environment coupling one obtains the time-independent Markovian decay rate 
\begin{equation}\label{gamma_M}
 \gamma_{\mathrm M} = \gamma(t\to\infty) = 2\pi J(\omega_0)
\end{equation}
and the time-independent renormalized frequency
\begin{equation} \label{def-omega_r-Markov}
 \omega^{\mathrm M}_r 
 = \omega_r(t\to\infty)
 = \omega_0 + \mathrm{P} \int_0^\infty \!\!\! d\omega \frac{J(\omega)}{\omega_0 - \omega},
\end{equation}
while $N(t)$ has to be replaced by the Planck distribution $N_0 = 1/(e^{\beta\omega_0}-1)$. These replacements 
directly lead to the well-known Born-Markov approximation for the quantum master equation in Lindblad form:
\begin{eqnarray}\label{meq-Born-Markov}
 \frac{d}{dt}\rho_S &=& -i\omega^{\mathrm M}_r[a^{\dagger}a,\rho_S] \nonumber \\
 && + \gamma_{\mathrm M} (N_0+1) \left[a\rho_Sa^{\dagger}-\frac{1}{2}\left\{a^{\dagger}a,\rho_S\right\}\right] \nonumber \\
 && + \gamma_{\mathrm M} N_0 \left[a^{\dagger}\rho_Sa-\frac{1}{2}\left\{aa^{\dagger},\rho_S\right\}\right].
\end{eqnarray}
As is well known \cite{Breuer2002} a renormalized frequency $\omega^{\mathrm M}_r$ even appears 
within the Born-Markov approximation, for which the frequency shift $\Delta\omega(t\to\infty)$ 
is given by the principal value integral in Eq.~\eqref{def-omega_r-Markov}. However, when applying
the master equation \eqref{meq-Born-Markov} e.g. in quantum optics the frequency shift is often very small for weak couplings
and, hence, is commonly neglected completely. This is consistent in the sense that
the stationary solution of the master equation \eqref{meq-Born-Markov}
describes a Boltzmann distribution over the energy levels of the bare
oscillator with frequency $\omega_0$. Therefore, in the following we define
results of the Born-Markov approximation as results which are obtained by modelling
the coupling to the baths by the master equation \eqref{meq-Born-Markov}
with $\omega^{\mathrm M}_r$ replaced by $\omega_0$.

\subsection{Otto cycle}\label{sec:otto-cycle}


We study a quantum Otto cycle using a harmonic oscillator with time dependent frequency
as working system. Our aim is to compare the results from the open system dynamics in the standard 
Born-Markov approximation with the exact approach in the non-Markovian regime. 

\subsubsection{Standard Otto cycle}

\begin{figure}[tb]
\includegraphics[width=0.99\linewidth]{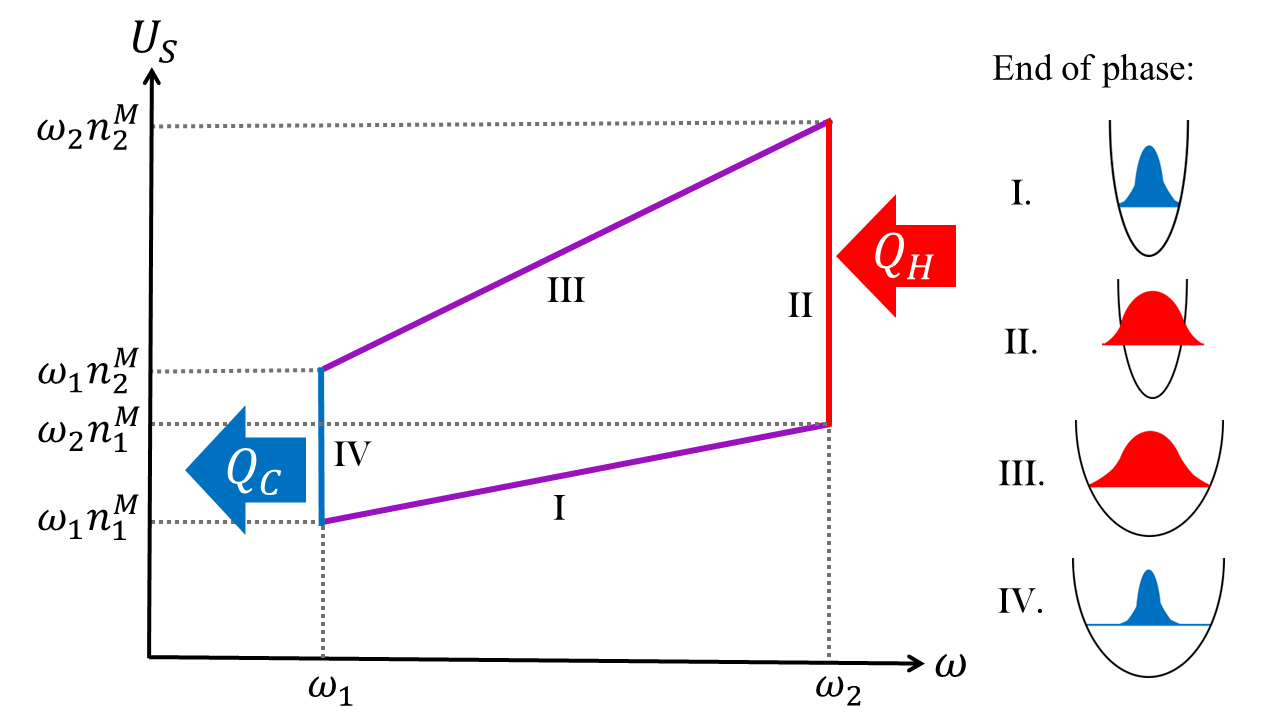}
\caption{Standard quantum Otto cycle of a harmonic oscillator. Phases I and III are adiabatic compression and expansion 
processes, respectively, while phases II and IV consist of isochoric heating and cooling by coupling to heat baths.}
\label{fig:Markov_cycle}
\end{figure}

The standard quantum Otto cycle \cite{Kosloff2017} consists of four phases
(see Fig.~\ref{fig:Markov_cycle}): 

\begin{enumerate}[I.]

\item Adiabatic compression: The frequency $\omega(t)$ of the harmonic oscillator is varied from the initial frequency
$\omega_1$ to the final frequency $\omega_2 > \omega_1$. During this phase the system is decoupled from
the baths such that the dynamics is unitary with Hamiltonian
\begin{eqnarray}\label{eq:driving_hamiltonian}
 H_S (t) = \omega(t) a^\dagger a.
\end{eqnarray}  
Consequently, there is no heat exchange, only work is performed on the system. For the Hamiltonian 
\eqref{eq:driving_hamiltonian} the mean occupation number $n=\langle a^\dagger a \rangle$ is obviously
a conserved quantity. At the initial time the oscillator  is assumed to be in a thermal equilibrium state and, hence,
we have 
\begin{equation}
 n_{1}^\mathrm{M} = \frac{1}{e^{\omega_1/k_\mathrm{B}T_1} - 1},
\end{equation}
where $T_1$ is the temperature of the cold bath and the index $\mathrm{M}$ 
serves to indicate that the quantity refers to the standard Born-Markov treatment. 
The work and heat transfer in phase I are thus given by
\begin{equation}\label{W/Q_I}
    W_\text{I} = (\omega_2 - \omega_1) n_1^\mathrm{M}, \qquad Q_\text{I} = 0.
\end{equation}
Note that we are assuming here that $a$ and $a^\dagger$ are frequency independent. When starting from the
expression $\frac{1}{2} p^2 + \frac{1}{2} \omega^2(t) x^2$ for the system Hamiltonian this assumption
corresponds to fully adiabatic motion during this phase, which means that the adiabaticity
parameter $Q^*$ is equal to 1 \cite{Deffner2008Feb,Husimi1953Apr}. We emphasize that this
assumption is made here only to simplify the presentation since our main goal is a comparison between
the standard Markovian and the non-Markovian treatment.

\item Isochoric heating: The oscillator is connected to a hot bath at temperature $T_2$ and evolves 
until thermalization is reached. The coupling to the bath is modelled through the Born-Markov master equation
\eqref{meq-Born-Markov} which implies that the mean occupation number in thermal equilibrium becomes
\begin{equation}
 n_{2}^\mathrm{M} = \frac{1}{e^{\omega_2/k_\mathrm{B}T_2} - 1}.
\end{equation}
Note that the steady state does not depend on the initial conditions, but only on the temperature of the bath $T_2$ 
and on the frequency $\omega_2$. The frequency remains constant at $\omega_2$, such that all changes of 
internal energy corresponds to heat exchange $Q_H$ and that there is no work contribution. Hence, we have
\begin{equation}\label{W/Q_II}
    W_\text{II} = 0, \qquad Q_H = Q_\text{II} = \omega_2(n^\mathrm{M}_2-n^\mathrm{M}_1).
\end{equation}

\item Adiabatic expansion: As in stroke I the oscillator is disconnected from the bath, and the frequency of the system 
is driven back from $\omega_2$ down to $\omega_1$ following a unitary dynamics given by the Hamiltonian
\eqref{eq:driving_hamiltonian}. The occupation number remains constant at the value $n_2^\mathrm{M}$. This leads to
\begin{equation}\label{W/Q_III}
    W_\text{III} = (\omega_1 - \omega_2) n_2^\mathrm{M}, \qquad Q_\text{III} = 0.
\end{equation}

\item Isochoric cooling: Finally, the oscillator is connected to a cold bath at temperature $T_1 < T_2$, and evolves at 
constant frequency $\omega_1$ until it thermalizes with mean occupation number $n_1^\mathrm{M}$. 
Hence, there is only heat exchange with the cold bath and no work contribution which yields:
\begin{equation}\label{W/Q_IV}
    W_\text{IV} = 0, \qquad Q_\text{IV} = \omega_1(n^\mathrm{M}_1-n^\mathrm{M}_2).
\end{equation}

\end{enumerate}

We define the efficiency $\eta$ of the cycle as
the ratio of the total work output $W_\mathrm{out}$
and the heat input $Q_H$ from the hot bath,
\begin{equation}\label{def-eta}
 \eta = \frac{W_\mathrm{out}}{Q_H},
\end{equation} 
when $W_\mathrm{out} > 0$, and as $0$ otherwise. 
According to Eqs.~\eqref{W/Q_I}-\eqref{W/Q_IV} in the Markovian case the total work output takes the form
\begin{equation}\label{eq:Wout_markov}
 W_\mathrm{out} = - (W_\text{I} + W_\text{III}) = (\omega_2 - \omega_1) (n_2^\mathrm{M} - n_1^\mathrm{M}),
\end{equation}
while the heat input $Q_H$  from the hot bath is given in Eq.~\eqref{W/Q_II}. Thus we obtain the well-known
expression for the efficiency of the cycle within the Born-Markov approximation \cite{Kosloff2017},
\begin{equation}\label{eq:eff_markov}
    \eta^\mathrm{M} = 1 - \frac{\omega_1}{\omega_2}.
\end{equation}
Employing Eq.~\eqref{eq:Wout_markov} we find that the condition on the frequencies and temperatures 
for the system to have net work output ($W_\mathrm{out} > 0$) and, thus, to operate as a heat engine, can be written as 
\begin{equation}\label{eq:condition_markov}
    \frac{T_1}{T_2} \omega_2 < \omega_1 < \omega_2.
\end{equation}

\subsubsection{Non-Markovian Otto cycle}

\begin{figure}[tb]
\includegraphics[width=0.99\linewidth]{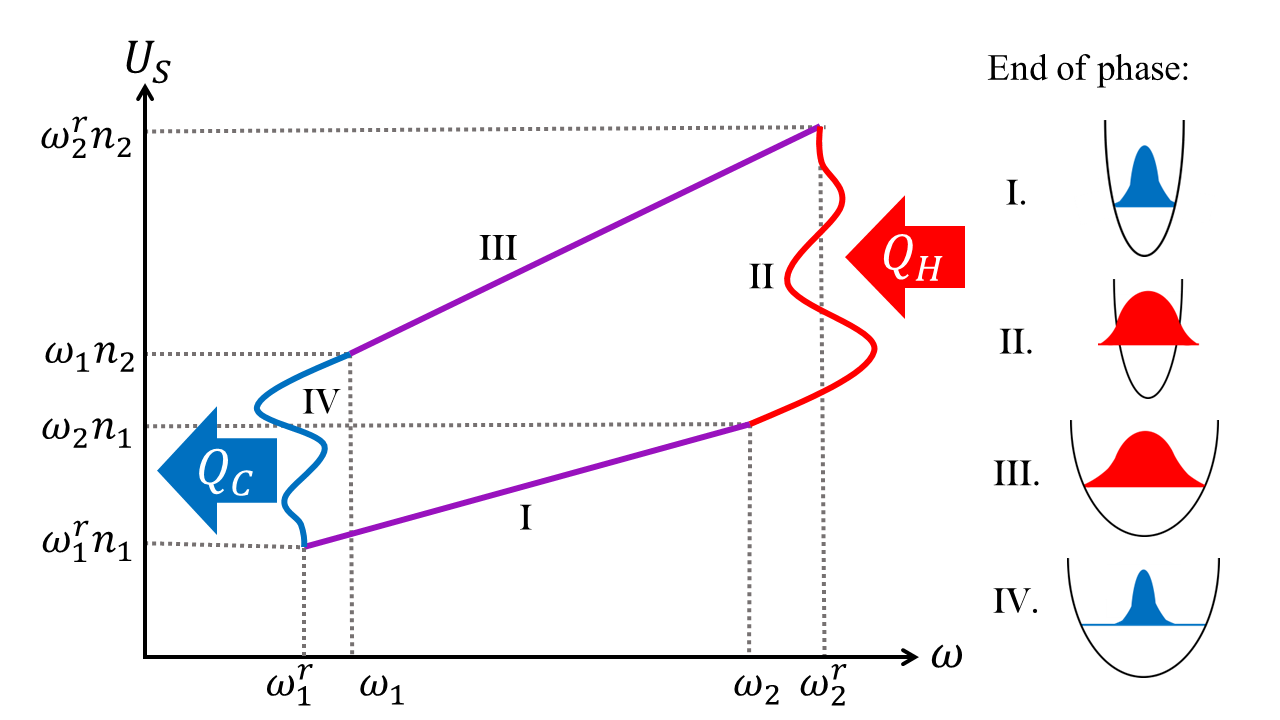}
\caption{Non-Markovian quantum Otto cycle of a harmonic oscillator. In this case, phases II and IV entail a time-dependent renormalization of the frequency due to the interaction with the bath.}
\label{fig:non-markov_cycle}
\end{figure}

In the following we will compare the exact approach, which is valid for arbitrary couplings and non-Markovian dynamics, 
with the standard treatment within the Born-Markovian approximation discussed in the previous section. 
The crucial differences to the standard approach are the time-dependent renormalization of the frequency of the system oscillator and the time dependence of the occupation number, as is visualized schematically in Fig.~\ref{fig:non-markov_cycle}.
Here we assume the non-Markovian bath to be such that the system still thermalizes to a unique occupation number 
$n_{1,2}$ depending on the temperature $T_{1,2}$ of the bath, and to a unique renormalized frequency 
$\omega_{1,2}^r$ depending on the bare frequency $\omega_{1,2}$ of the oscillator. 
The steady state also depends on the coupling 
and on the structure of the bath. The strokes of the cycle are then altered in the following way.

\begin{enumerate}[I.]

\item Adiabatic compression: This phase starts now at frequency $\omega_1^r$ and occupation number $n_1$
which remains constant throughout the increase of the frequency up to $\omega_2$. Since the baths are decoupled
we have again unitary evolution and only a work contribution which is given by
\begin{equation}
 W_\text{I} = \left( \omega_2 - \omega_1^r \right) n_1.
\end{equation}
Initially, the oscillator is assumed to be in a thermal equilibrium state at the temperature $T_1$ of the cold bath, 
i.e. we have
\begin{equation}\label{n_1}
 n_1 = n_1(t\to\infty) = \langle a^\dagger a \rangle_{t \rightarrow \infty}^{T_1}.
\end{equation}

\item Heating phase: The oscillator is coupled to the hot bath with temperature $T_2$. The
exact dynamics of the oscillator is given by the non-Markovian master equation \eqref{tcl-meq}
involving the effective Hamiltonian \eqref{K_S-fano-anderson} and the dissipator \eqref{dissipator-fano-anderson},
giving rise to a time dependent oscillator frequency, which results in a work contribution to the energy change. 
Hence, this is no longer an isochoric process, as there is generally both work and heat exchange between the system and 
the bath due to the non-Markovian behaviour. The final frequency is $\omega_2^r=\omega_2^r(t\to\infty)$ and the final 
occupation number becomes
\begin{equation}\label{n_2}
 n_2 = n_2(t\to\infty) = \langle a^\dagger a \rangle_{t \rightarrow \infty}^{T_2}.
\end{equation}
Note that these quantities are generally different from $\omega_2$ and $n_2^M$, respectively. Using 
Eqs.~\eqref{work} and the effective Hamiltonian \eqref{K_S-fano-anderson} we find the work contribution
\begin{equation}
 W_\text{II} = \int_0^\infty dt \; \dot{\omega}_2^r (t) n_2 (t),    \label{eq:W_II}
\end{equation}
where the expressions for $\omega_2^r(t)$ and $n_2(t) = \langle a^\dagger a \rangle^{T_2}_t$ are found in 
Eqs.~\eqref{def-omega_r} and \eqref{moment-a*a}. 

\item Adiabatic expansion: The oscillator is again disconnected from the baths, and the frequency is driven down to 
$\omega_1$, while the occupation number $n_2$ stays constant. The change of internal energy is fully due to work 
exchange which is given by
\begin{equation}
 W_\text{III} = \left( \omega_1 - \omega_2^r \right) n_2
\end{equation}
with $n_2$ given by Eq.~\eqref{n_2}.

\item Cooling phase: The oscillator is coupled to the cold heat bath with temperature $T_1$ which leads to relaxation
to thermal equilibrium with frequency $\omega_1^r=\omega_1^r(t\to\infty)$ and occupation number $n_1$ given by 
Eq.~\eqref{n_1}. As in phase II we have both heat and work exchange, the latter given by
\begin{equation}
 W_\text{IV} = \int_0^\infty dt \; \dot{\omega}_1^r (t) n_1 (t).    \label{eq:W_IV}
\end{equation}

\end{enumerate} 

The efficiency is again defined by \eqref{def-eta}. However, for the non-Markovian treatment the total work output 
consists of contributions from all four phases of the cycle: 
\begin{equation}
 W_\mathrm{out} = - (W_\text{I} + W_\text{II} + W_\text{III} + W_\text{IV}). 
\end{equation}
The expression for the heat input from the hot bath is given by
\begin{equation}
    Q_H = \int_0^\infty dt \; \omega_2^r (t) \dot{n}_2 (t) ,
    \label{eq:QH_FA}
\end{equation}
but we can also make use of the first law of thermodynamics to determine $Q_H$ by means of
\begin{equation}
 Q_H = \Delta U_S^\text{II} - W_\text{II} = \omega_2^r n_2 - \omega_2 n_1 - W_\text{II}.
\end{equation}

In our numerical simulations we consider two baths at temperatures $T_1$ and $T_2$ with identical
structured spectral densities described by a Lorentzian
\begin{equation}\label{eq:lorentzian_spectral_density_wc}
    J(\omega) = \frac{\gamma_0}{2 \pi} \frac{\lambda^2}{(\omega_c - \omega)^2 + \lambda^2},
\end{equation}
where $\lambda$ represents the spectral width and $\omega_c=\omega_0-\Delta$ describes the position 
of the peak with $\omega_0$ the bare frequency of the oscillator and $\Delta$ the detuning. We note that we
have modified the Lorentzian shape of the spectral density in the regime of very low frequency
modes in order for the noise integral \eqref{noise-integral} to be finite; see appendix \ref{app-B} for details.
Depending on the phase of the cycle we have $\omega_0=\omega_{1,2}$ and, considering a fixed $\omega_c$,
the detuning is $\Delta=\Delta_2=\omega_2-\omega_c$ for the heating phase, and
$\Delta=\Delta_1=\omega_1-\omega_c$ for the cooling phase.
Using Eq.~\eqref{gamma_M} we see that the parameter $\gamma_0$, which is used as a measure for the
strength of the system-bath coupling, corresponds to the Markovian decay rate at resonance ($\Delta=0$).

With the spectral density \eqref{eq:lorentzian_spectral_density_wc} Eq.~\eqref{green-function} can easily be solved by Laplace transformation if one extends the range of the integral \eqref{memory-kernel-2} to the entire real axis, leading to the solution
\begin{eqnarray}\label{eq:G_lorentzian}
    G(t) = \frac{e^{-i \omega_0 t}}{\mu_2 - \mu_1} \left(\mu_2 e^{\mu_1 t} - \mu_1 e^{\mu_2 t}\right),
\end{eqnarray}
where $\mu_{1,2}$ are the roots of the quadratic equation 
\begin{equation}
 \mu^2 + (\lambda - i \Delta) \mu + \frac{\gamma_0 \lambda}{2} = 0.
\end{equation}
For this particular case, the coupling to the bath induces a time-dependent renormalization 
of the oscillator frequency which, according to Eq.~\eqref{def-omega_r}, is given by 
\begin{equation}\label{eq:wr_lorentzian}
 \omega_r(t) 
 = \omega_0 - \Im \left( \mu_1 \mu_2 \frac{e^{\mu_1 t} - e^{\mu_2 t}}{\mu_2 e^{\mu_1 t} - \mu_1 e^{\mu_2 t}} \right).
\end{equation}
Using Eq.~\eqref{eq:G_lorentzian} one can show that in this model the system always relaxes to a unique 
steady state representing a thermal equilibrium state. Moreover, if we take the long time limit of 
\eqref{eq:wr_lorentzian} we find that  
$\omega_r (t \rightarrow \infty) > \omega_0 $ for $\Delta > 0$ and $\omega_r (t \rightarrow \infty) < \omega_0$ 
for $\Delta < 0$. This behavior can be seen explicitly for the Born approximation of the renormalized frequency which is given by
\begin{eqnarray}\label{eq:wr_Born}
    \omega_r (t) &=& \omega_0 + \frac{\gamma_0 \lambda \Delta / 2}{\lambda^2 + \Delta^2} \left[1 - e^{- \lambda t} \Big( \cos \Delta t + \frac{\lambda}{\Delta} \sin \Delta t \Big) \right] \nonumber \\
    &\to& \omega_0 + \frac{\gamma_0 \lambda \Delta / 2}{\lambda^2 + \Delta^2} \qquad (t \to \infty),
\end{eqnarray}
showing that the sign of the frequency shift is given by the sign of
$\Delta$.

Recently, the impact of the switching on and off of the coupling to the baths has been studied in detail investigating 
the system-bath interaction energy \cite{Shirai2021,Latune2023}. We note that in our approach this effect
is taken into account through the time dependence of the renormalized frequency of the system oscillator.

\section{Simulation results}\label{sec:numerics}

\begin{figure}[tb]
\includegraphics[width=0.99\linewidth]{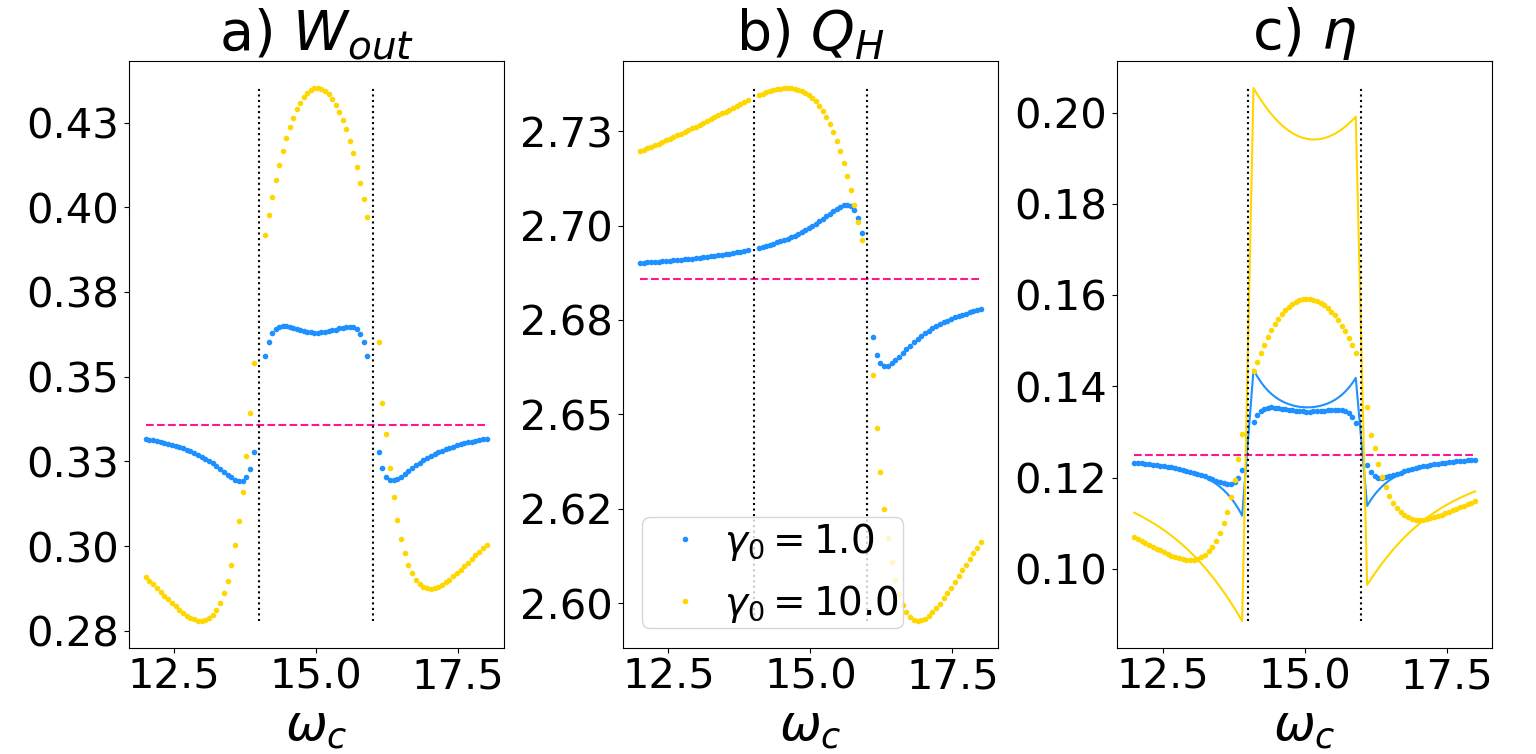}
\caption{a) Total work output, b) net heat input in phase II and c) efficiency as a function of the central frequency in the spectral density, for different values of $\gamma_0$. The dots are calculated through the exact approach, the pink dashed lines indicate the results from the Born-Markov approximation and the vertical black dotted lines indicate where $\omega_c = \omega_{1,2}$. The solid line in the efficiency graph shows the renormalized efficiency \eqref{eq:eff_freq}. The parameters used are $\omega_1 = 14.0$, $\omega_2 = 16.0$, $k_{\mathrm B}T_1 = 15.0$, $k_{\mathrm B}T_2 = 20.0$ and $\lambda = 0.2$.}
\label{fig:vs_wc_diff_gamma}
\end{figure}

\begin{figure}[tb]
\includegraphics[width=0.99\linewidth]{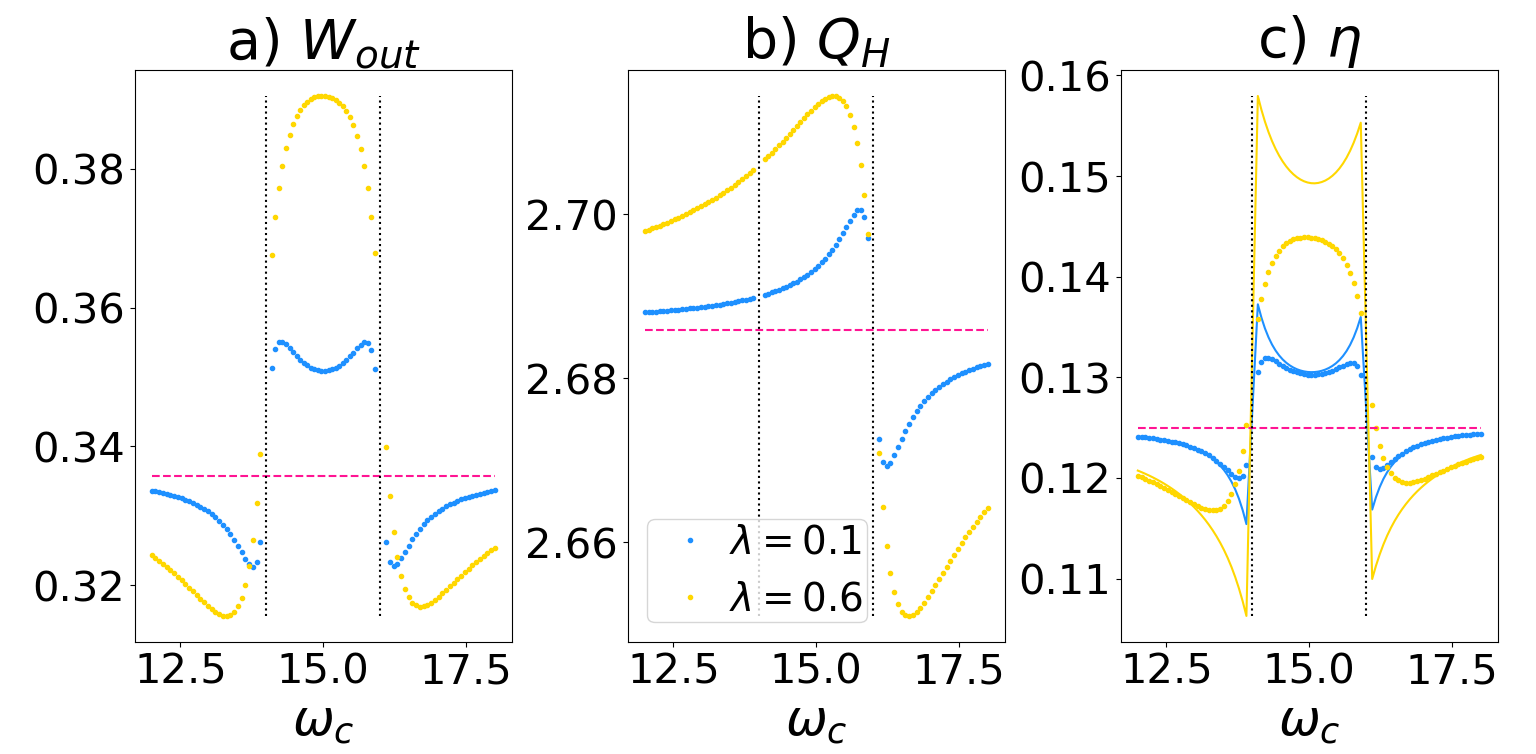}
\caption{a) Total work output, b) net heat input in phase II and c) efficiency as a function of the central frequency in the spectral density, for different values of $\lambda$. The points are calculated through the exact approach, the pink dashed lines indicate the results from the Born-Markov approximation and the vertical black dotted lines indicate where $\omega_c = \omega_{1,2}$. The solid line in the efficiency graph shows the renormalized efficiency \eqref{eq:eff_freq}. The parameters used are $\omega_1 = 14.0$, $\omega_2 = 16.0$, $k_{\mathrm B}T_1 = 15.0$, $k_{\mathrm B}T_2 = 20.0$ and $\gamma_0 = 1.0$.}
\label{fig:vs_wc_diff_lambda}
\end{figure}

Let us now discuss some numerical simulations of important quantities characterizing the performance of the
Otto cycle. All quantities and parameters presented in the following are either dimensionless or have the 
dimension of energy (or frequency, since we have set $\hbar=1$). Therefore, all dimensionful quantities will be
represented in terms of an arbitrary energy or frequency unit.

In Fig.~\ref{fig:vs_wc_diff_gamma} we show the behaviour of a) the total work output $W_{out}$, b) the net heat input from the hot bath $Q_H$ and c) the efficiency $\eta$ when varying $\omega_c$ (the position of the peak of the spectral densities of both the hot and the cold bath). The structure of the spectral density does not play a role in the Born-Markov approximation, therefore this parameter will only affect the exact dynamics. The pink dashed lines show the Born-Markov case for comparison, and we see that they remain constant for all $\omega_c$. The dots show the results from the exact dynamics for different values of $\gamma_0$, for the weak coupling case (blue) and the strong coupling case (yellow). 
As expected, we observe that in the weak coupling regime the results are generally closer to the Born-Markov approximation than in the strong coupling regime.

We also look at the behaviour of work output, heat input and efficiency with varying $\omega_c$ for different values of the spectral width $\lambda$ in Fig.~\ref{fig:vs_wc_diff_lambda}. As can be seen from the second-order Born approximation in Equation \eqref{eq:wr_Born}, the value of $\lambda$ also affects the coupling strength. This fact is reflected in the graph: For a narrow spectral density (blue lines) the results come closer to those from the Born-Markov approximation than for a wider one (yellow lines).

We note that in both cases shown in Figs.~\ref{fig:vs_wc_diff_gamma} and \ref{fig:vs_wc_diff_lambda} 
when changing $\omega_c$ the work output deviates more strongly from
the Born-Markov approximation than the heat input. 
In our numerical simulations we find this to be a general behaviour when varying parameters of the spectral density. Therefore, the behaviour of work with changing characteristics of the spectral density has a greater influence on the behaviour of the efficiency than that of heat. 

\begin{figure}[tb]
\includegraphics[width=0.99\linewidth]{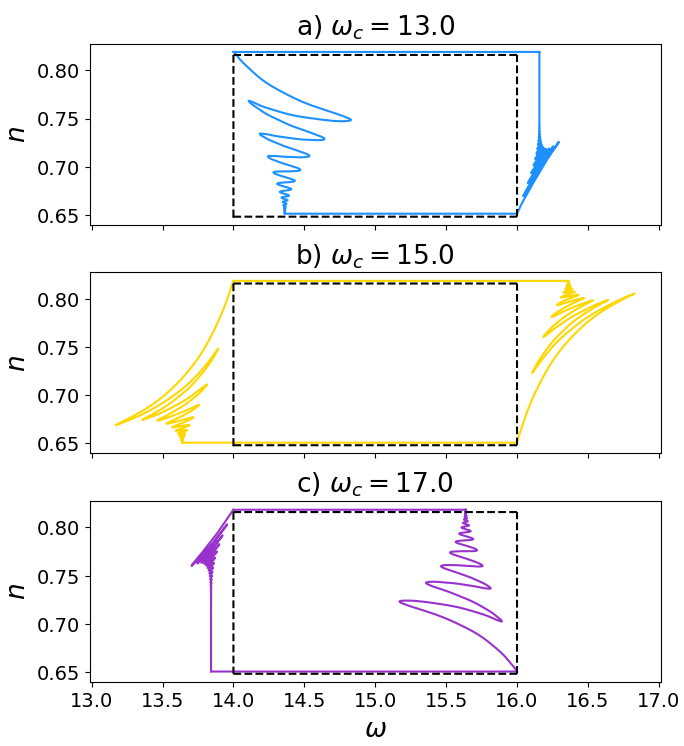}
\caption{The Otto cycle represented in the 
$(\omega,n)$-plane for different values of $\omega_c$. The dashed black lines show the Markovian Otto cycle, while the colored solid lines stand for the exact results. The area enclosed by the curves gives the total work output. The parameters used are $\omega_1 = 14.0$, $\omega_2 = 16.0$, $k_{\mathrm B}T_1 = 15.0$, $k_{\mathrm B}T_2 = 20.0$, $\gamma_0 = 5.0$ and $\lambda = 0.2$. We see that the area is bigger than in the Born-Markov approximation for the case $\omega_1 < \omega_c < \omega_2$ (b), and smaller otherwise (a and c).}
\label{fig:cycle_wc}
\end{figure}

Figures~\ref{fig:vs_wc_diff_gamma} and \ref{fig:vs_wc_diff_lambda}
clearly show a characteristic behaviour: Work output and efficiency are greater than in the Born-Markov case in the region $\omega_1 < \omega_c < \omega_2$, and smaller outside this region. The first intuition for this fact comes from Fig.~\ref{fig:cycle_wc}, where we represent, for different values of $\omega_c$, the Otto cycle in the
$(\omega,n)$-plane, where $\omega$ is the frequency of the oscillator
and $n$ the mean occupation number.
The colored solid curves represent the exact cycle, while the dashed black lines show the standard Born-Markov approximation. The area enclosed by the curves corresponds to the total work output. We can already see in this picture that the work output is higher for the case $\omega_1 < \omega_c < \omega_2$ than in the Born-Markov approximation (Fig.~\ref{fig:cycle_wc}b). This case corresponds to $\Delta_2 > 0$ and $\Delta_1 < 0$ and, consequently, to $\omega_1^r (t \rightarrow \infty) < \omega_1$ and $\omega_2^r (t \rightarrow \infty) > \omega_2$, which results in the cycle taking place within a larger range of frequencies. For the cases $\omega_1 > \omega_c$ (a) and $\omega_2 < \omega_c$ (c) we see that the work output is smaller than in the standard approach, and that here the frequency renormalization reduces the work output, instead of enhancing it. 

This behaviour of the work output translates into efficiency, which is
also larger than the Born-Markov efficiency within the region 
$\omega_1 < \omega_c < \omega_2$. A simple approximation can be derived if we assume that within second order 
$\omega_r (t)$ relaxes much more rapidly than $n(t)$. We see from \eqref{eq:wr_Born} that $\omega_r(t)$ relaxes on 
the time scale $\lambda^{-1}$, while $n(t)$ relaxes on the time scale $\gamma_M^{-1}$, where $\gamma_M$ is the
Markovian relaxation rate given by Eq.~\eqref{gamma_M}, which yields in the present case 
$\gamma_M=\gamma_0\lambda^2/(\lambda^2+\Delta^2)$. Thus, the assumption $\lambda^{-1} \ll \gamma_M^{-1}$
leads to the condition 
\begin{equation} \label{cond-ren-eff}
 \frac{\gamma_0\lambda}{\lambda^2+\Delta^2} \ll 1.
\end{equation}
Under this condition we
can use \eqref{eq:W_II} and
\eqref{eq:W_IV} to get the approximate expressions
\begin{equation}\label{eq:W_approx}
    W_\text{II} \approx n_1 (\omega_2^r - \omega_2), \qquad W_\text{IV} \approx n_2 (\omega_1^r - \omega_1),
\end{equation}
such that the efficiency of the cycle may be approximated by the
renormalized effciency $\eta_r$:
\begin{equation}\label{eq:eff_freq}
 \eta \approx \eta_r = 1 - \frac{\omega_1^r}{\omega_2^r}.
\end{equation}
Note that $\eta_r$ is obtained from the Markovian
efficiency \eqref{eq:eff_markov} by replacing the bare frequencies with the renormalized frequencies.
The renormalized efficiency \eqref{eq:eff_freq} is represented by the solid lines in Figs.~\ref{fig:vs_wc_diff_gamma}c 
and \ref{fig:vs_wc_diff_lambda}c. For the blue curves condition \eqref{cond-ren-eff} is satisfied outside the
resonances at $\omega_c=\omega_{1,2}$, and we see that \eqref{eq:eff_freq} indeed provides a good
approximation for the cycle efficiency. Note that in the vicinity of the resonances ($\Delta \approx 0$) condition
\eqref{cond-ren-eff} is violated which explains why the approximation \eqref{eq:eff_freq} is less accurate
close to resonance.

\begin{figure}[tb]
\includegraphics[width=0.99\linewidth]{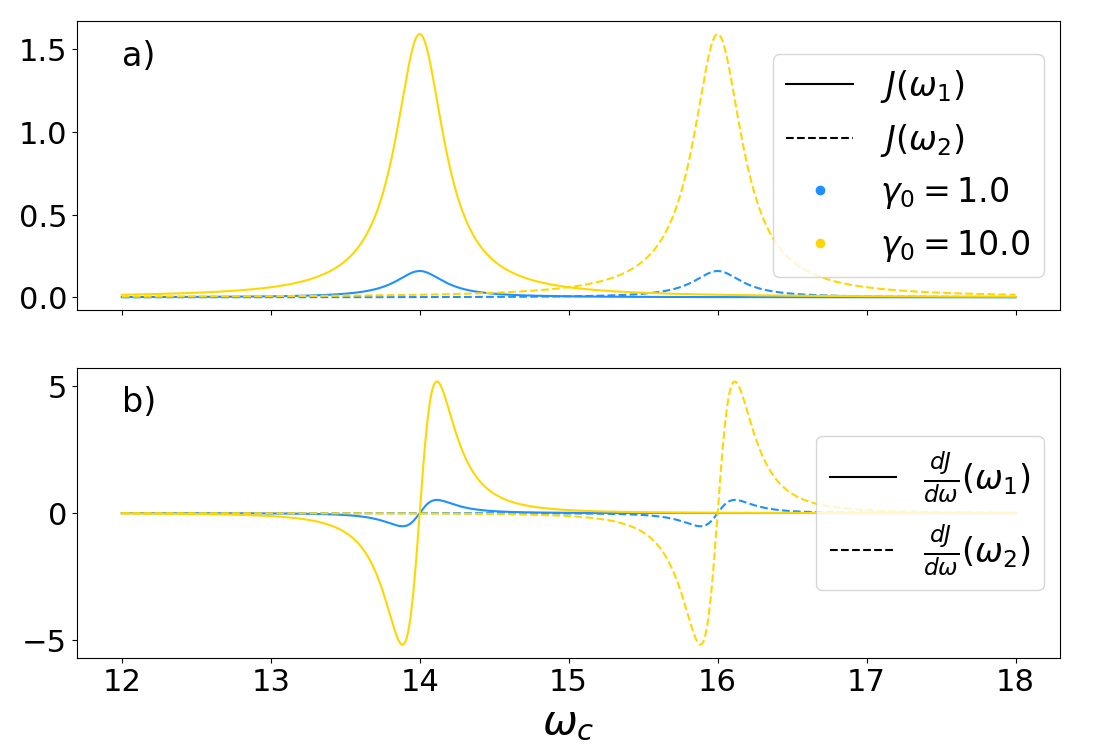}
\caption{a) Value of the spectral density and b) slope of the spectral density at the initial frequencies of the central harmonic oscillator: $J(\omega_1)$ (solid line) and $J(\omega_2)$ (dashed line) as a function of $\omega_c$, for different values of $\gamma_0$. The frequencies, temperatures and spectral width $\lambda$ are the same as in Fig.~\ref{fig:vs_wc_diff_gamma}.}
\label{fig:spectral_density_gamma}
\end{figure}

\begin{figure}[tb]
\includegraphics[width=0.99\linewidth]{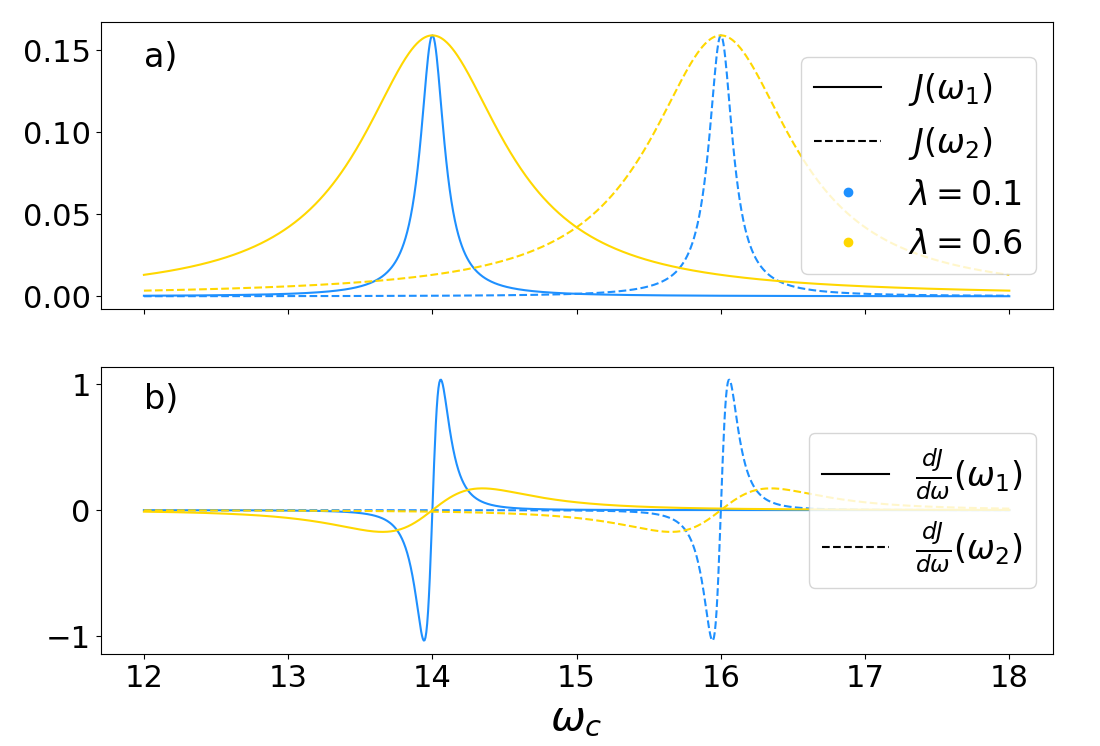}
\caption{a) Value of the spectral density and b) slope of the spectral density at the initial frequencies of the central harmonic oscillator: $J(\omega_1)$ (solid line) and $J(\omega_2)$ (dashed line) as a function of $\omega_c$, for different values of $\lambda$. The frequencies, temperatures and coupling strength $\gamma_0$ are the same as in Fig.~\ref{fig:vs_wc_diff_lambda}.}
\label{fig:spectral_density_lambda}
\end{figure}

To further understand the curves shown in Figs.~\ref{fig:vs_wc_diff_gamma} and \ref{fig:vs_wc_diff_lambda} we give an interpretation based on the height and the slope of the spectral density at the initial frequencies of the harmonic oscillator when coupled to each bath. 
In Figs.~\ref{fig:spectral_density_gamma} and \ref{fig:spectral_density_lambda} we represented the height (top) and the slope (bottom) of the spectral density at $\omega = \omega_1$ (solid lines) and $\omega = \omega_2$ (dashed lines), i.e., $J(\omega_{1,2})$ and  $\frac{d}{d \omega} J (\omega_{1,2})$, with varying $\omega_c$ on the horizontal axis. 

First, we will analyze the different behaviour of work output, heat input and efficiency for different values of the coupling strength $\gamma_0$. In Fig.~\ref{fig:spectral_density_gamma} the lines show $\gamma_0 = 1.0$ (blue) and $\gamma_0 = 10.0$ (yellow), and all other parameters are the same as in Fig.~\ref{fig:vs_wc_diff_gamma}. When we look at the curves in Fig.~\ref{fig:vs_wc_diff_gamma}, we find, first of all, that the closer to resonance with the corresponding baths ($\omega_c \rightarrow \omega_{1,2}$), the more the results deviate from the standard Born-Markov approximation (except at exact resonance, see below), while far away from resonance the approximation predicts the results accurately. We can relate this to the fact that the height and the slope of the spectral density at $\omega_{1,2}$ close to resonance take significant values, while far away from resonance they have decayed to negligible quantities, and the bath looks flat in the vicinity of the initial frequency of the central harmonic oscillator. 
Additionally, we see that, for weak coupling (blue lines), $W_{out}$ presents a dip in the region $\omega_1 < \omega_c < \omega_2$, while for strong coupling (yellow lines) work output and efficiency have no inflection point in this region. If we look at the spectral densities for these two values of $\gamma_0$, we see that in the weak coupling regime the spectral densities and their slopes have decayed to negligible values for $\omega_c$ centrally located between $\omega_1$ and $\omega_2$, while for strong coupling the spectral densities still take high values in this region, with significant slopes, and therefore strong effects in both the heating and the cooling phases combine. This is the reason why work output (and consequently efficiency) is maximally enhanced in this region for strong coupling.
We also see that at resonance with either of the baths, the quantities seem to come close again to the Born-Markov approximation. If we look at the slope of the spectral densities in Fig.~\ref{fig:spectral_density_gamma}, we see that it vanishes as it approaches resonance ($\omega_c = \omega_{1,2}$), resembling a flat bath.

In Fig.~\ref{fig:spectral_density_lambda} the height (a) and slopes (b) of the spectral density at $\omega_{1,2}$ can be found for different values of $\lambda$, corresponding to the results shown in Fig.~\ref{fig:vs_wc_diff_lambda}. The faster decay of the spectral density and its slope for small $\lambda$ (blue) can account for the higher resemblance of the results to those in the Born-Markov approximation far from resonance. On the contrary, for large $\lambda$ (yellow) the cycle is most altered for intermediate $\omega_c$ between $\omega_1$ and $\omega_2$, when the height and slope of both $J(\omega_1)$ and $J(\omega_2)$ are still significant, meaning that both the heating and the cooling phases feel the effects of a strongly coupled, structured bath.

\begin{figure}[tb]
\includegraphics[width=0.99\linewidth]{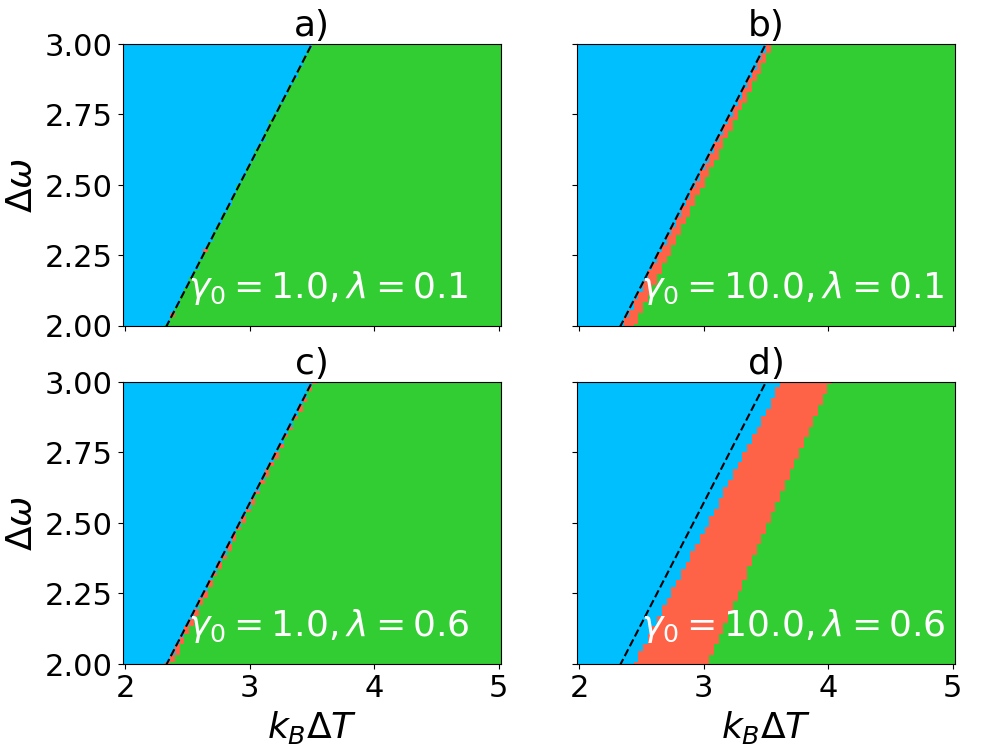}
\caption{Representation of the parameter regions for different behaviours of the cycle: In the green region the cycle behaves as a heat engine ($W_{out} > 0$ and $Q_H > 0$), in the blue region as a refrigerator ($W_{out} < 0$ and $Q_H < 0$) and in the red region as a heater ($W_{out} < 0$ and $Q_H > 0$). The black dashed lines represent the limit between the heat engine and the refrigerator regions in the standard Born-Markov approximation (the cycle cannot function as a heater in this case). We choose $\omega_c = 15.0$ as a reference frequency and $k_B T_0 = 17.5$ as a reference temperature, and we vary $\omega_{1,2}$ and $k_B T_{1,2}$ symmetrically (such that $\omega_{1,2} = \omega_c \mp \Delta \omega / 2$ and $k_B T_{1,2} = k_B T_0 \mp k_B \Delta T / 2$).}
\label{fig:functioning_regions}
\end{figure}

We now turn to look at the different modes of operation of the cycle in different parameter ranges. In the standard Born-Markov approximation this distinction is clear: The cycle operates as a heat engine ($W_{out} > 0$ and $Q_H > 0$) for the configurations of frequencies and temperatures that satisfy \eqref{eq:condition_markov}, and as a refrigerator ($W_{out} < 0$ and $Q_H < 0$) otherwise (due to the fact that work output and heat input change sign at the same points). In the exact approach we can only determine the behaviour of the cycle numerically, and it will additionally depend on the specific parameters of the spectral density. We find that the exact approach can give rise to three different types of behaviour: As a heat engine or refrigerator as before, but also as a heater ($W_{out} < 0$ and $Q_H > 0$), where the cycle just uses work to amplify the flow of heat from the hot to the cold reservoir. 

In Fig.~\ref{fig:functioning_regions} we depict the regions in which the cycle operates as a heat engine (green), as a refrigerator (blue) or as a heater (red). The dashed black lines represent the limit between the heat engine and the refrigerator regions in the standard Born-Markov approximation. On the horizontal axis we change $k_B \Delta T$, where we have chosen a reference temperature $k_B T_0$ and vary the temperatures of the cold and the hot bath symmetrically such that $k_B T_{1,2} = k_B T_0 \mp k_B \Delta T / 2$. On the vertical axis we change $\Delta \omega$, where we have chosen a $\omega_c$ as the reference frequency and vary $\omega_{1,2} = \omega_c \mp \Delta \omega / 2$. It can be observed that in the exact approach the region in which the cycle operates as a heat engine is limited compared to the Born-Markov case, with work output being positive for a smaller range of frequencies and temperatures. For small $\gamma_0$ and $\lambda$ this difference is almost negligible, but it increases as we increase $\gamma_0$  and $\lambda$. Additionally, a new type of behaviour appears and expands as $\gamma_0$ and $\lambda$ increase: The cycle operates as a heater, where a work input is used to dissipate heat from the hot reservoir into the cold reservoir. In spite of this, for large values of $\gamma_0$ and $\lambda$ (Fig.~\ref{fig:functioning_regions}d) the refrigerator region broadens.

\begin{figure}[tb]
\includegraphics[width=0.99\linewidth]{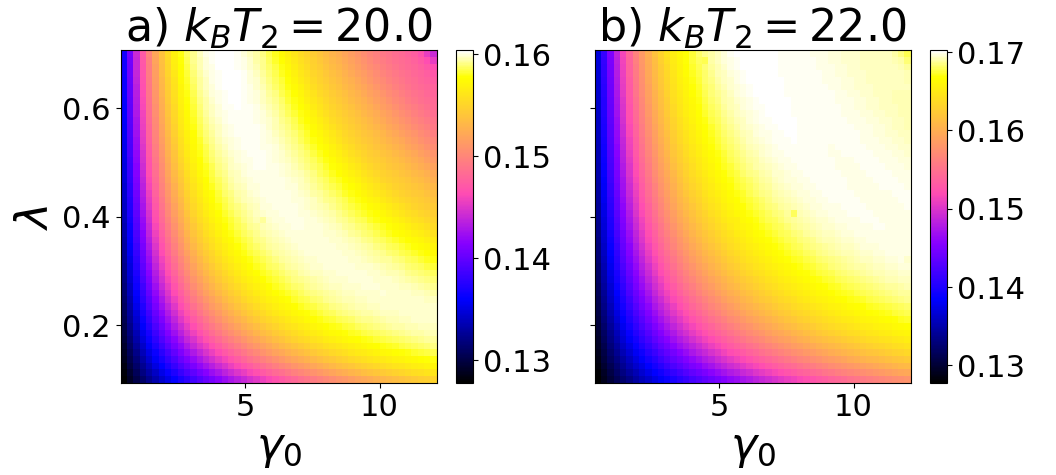}
\caption{Efficiency of the Otto cycle as a function of $\gamma_0$ on the horizontal axis and $\lambda$ on the vertical axis, for different values of the temperature of the hot reservoir: a) $k_{\mathrm B}T_2 = 20.0$ and b) $k_{\mathrm B}T_2 = 22.0$. The other parameters are: $\omega_1 = 14.0$, $\omega_2 = 16.0$, $\omega_c = 15.0$ and $k_{\mathrm B}T_1 = 15.0$.}
\label{fig:optimum_gamma_lambda}
\end{figure}

Finally, we analyze the values of $\gamma_0$ and $\lambda$ for which we can obtain the highest efficiencies. From Figs.~\ref{fig:vs_wc_diff_gamma} and \ref{fig:vs_wc_diff_lambda} it may seem that, for a spectral density satisfying $\omega_1 < \omega_c < \omega_2$, the higher the parameters $\gamma_0$ or $\lambda$ are, the higher is the efficiency. However, in Fig.~\ref{fig:optimum_gamma_lambda} we show the efficiency as a function of both $\gamma_0$ and $\lambda$ for a spectral density centered between $\omega_1$ and $\omega_2$, and we see that the behaviour of $\eta$ is not monotonic with increasing $\gamma_0$ or $\lambda$. We observe that, fixing one parameter, there exists an optimal value of the other parameter such that the efficiency is maximal. We show the results for two different temperatures of the hot bath $k_B T_2$, and it can be seen that the values of $\gamma_0$ and $\lambda$ for which the efficiency is maximal change with temperature. Therefore, stronger coupling can enhance the efficiency of the cycle, but only up to a limit, at which point it starts being a hindrance. Larger spectral width $\lambda$ can also start improving the efficiency, as the spectral density gets wider and starts having relevant effects on both the heating and cooling phases. However, if $\lambda$ becomes too large, the slope of the spectral density decreases, resembling a flat, Markovian bath and resulting in a lower efficiency for the cycle.

\section{Conclusions}\label{Conclu}
We have investigated the thermodynamics of the quantum Otto cycle in the presence of non-Markovian effects and strong system-bath coupling. Our analysis was based on a novel approach to quantum thermodynamics, which allowed us to exactly resolve the consequences of coupling a system to an arbitrary reservoir with memory effects. By comparing the standard Markovian Otto cycle with the exact non-Markovian treatment, we have identified feasible conditions under which the cycle exhibits enhanced work output and efficiency. These enhancements occur when the spectral density of the bath peaks at a frequency which is located within the driving range of the system, leading to an effective broadening of the range of frequencies in which the cycle operates. Conversely, outside these regions, the cycle performance either is lower or remains close to the standard Born-Markov approximation. When taking baths with a spectral density that enhances the efficiency, we encounter a small trade-off in the parameter region of frequencies and temperatures for which the cycle is producing work output, but this effect becomes noticeable only at very large coupling parameters. We have also shown that the behaviour of work output and efficiency strongly depends on the strength of the system-bath coupling and the width of the spectral density, showcasing how much the nuances of the bath spectral density can influence the performance of the cycle. 

It is important to acknowledge that our analysis focused solely on the quantum Otto cycle. Further research is required to investigate the implications of memory effects in other quantum thermal machine models and explore their influence on diverse thermodynamic processes. Still, the present study highlights the role of memory effects and structured environments as a promising avenue for the enhancement of current quantum thermal machines -- one notable implication of our findings being the potential for exploiting strong system-bath coupling to achieve more efficient quantum heat engines. 

While in this work we have not concerned ourselves with the duration of the cycle -- allowing fully adiabatic driving and infinitely long thermalization times -- this could in principle represent another positive aspect of considering reservoirs outside of the weak-coupling regimes. A promising path would be to investigate how the coupling strength can influence thermalization rates, possibly leading to shorter relaxation times and, in turn, higher efficiencies at maximum power. Moreover, while we have demonstrated enhancements in performance for specific scenarios, it still warrants exploration whether surpassing the Carnot efficiency is feasible with non-Markovian baths.

\acknowledgments

This project has received funding from the European Union's Framework Programme 
for Research and Innovation Horizon 2020 (2014-2020) under the 
Marie Sk\l{}odowska-Curie Grant Agreement No.~847471. The work has also been supported 
by the German Research Foundation (DFG) through FOR 5099.

\appendix

\section{Derivation of the exact master equation}\label{app-A}

We present some details on to derivation of the time-convolutionless exact master equation of the form \eqref{tcl-meq}
with effective Hamiltonian \eqref{K_S-fano-anderson} and dissipator \eqref{dissipator-fano-anderson}
for the microscopic model discussed in Sec.~\ref{sec:micro-model}. This master equation has been obtained
in \cite{Tu2008,Jin2010,Lei2012} using path integral techniques.
Here, we show how to derive it directly from the exact
expressions \eqref{moment-a}-\eqref{moment-a*a} for the first and second moments. To this end, we first
take the time derivatives to find the following exact equations of motion for these moments:
\begin{eqnarray} 
 \frac{d}{dt} \langle a \rangle_t &=& \frac{\dot{G}(t)}{G(t)} \langle a \rangle_t, \label{d-moment-a} \\
 \frac{d}{dt} \langle aa \rangle_t &=& 2 \frac{\dot{G}(t)}{G(t)} \langle aa \rangle_t, \label{d-moment-aa} \\
 \frac{d}{dt} \langle a^{\dagger}a \rangle_t &=&  2 \Re \left(\frac{\dot{G}(t)}{G(t)}\right) \langle a^{\dagger}a \rangle_t
 \nonumber \\
 && - 2 \Re \left(\frac{\dot{G}(t)}{G(t)}\right) I(t) + \dot{I}(t). \label{d-moment-a*a}
\end{eqnarray}
Since the microscopic Hamiltonian is quadratic we know that the dynamics preserves the Gaussianity
of states and, hence, that the master equation must be quadratic in the creation and annihilation operators
$a^\dagger$, $a$ of the central oscillator. The strategy is thus to write a suitable quadratic ansatz for the generator of
the master equation and to chose the coefficients in such a way that the moment equations obtained from
the master equation coincide with Eqs.~\eqref{d-moment-a}-\eqref{d-moment-a*a}. If we have found such an ansatz,
we can conclude that it yields the correct exact generator of the master equation.

Our ansatz for the master equation takes the form 
\begin{equation} \label{ansatz-tcl-meq}
 \frac{d}{dt}\rho_S(t) = -i \left[\tilde{K}_S(t),\rho_S(t)\right] + \tilde{\mathcal{D}}_t  [\rho_S(t)],
\end{equation}
where (compare to Eqs.~\eqref{K_S-fano-anderson} and \eqref{dissipator-fano-anderson})
\begin{equation} \label{ansatz-K_S-fano-anderson}
 \tilde{K}_S(t) = \Omega(t) a^{\dagger}a + ig(t)a^\dagger - ig^*(t)a,
\end{equation}
and
\begin{eqnarray} \label{ansatz-dissipator-fano-anderson}
  \tilde{\mathcal{D}}_t[\rho_S] &=&  
 d_1(t) \left[a\rho_Sa^{\dagger}-\frac{1}{2}\left\{a^{\dagger}a,\rho_S\right\}\right] \nonumber \\
 &+& d_2(t) \left[a^{\dagger}\rho_Sa-\frac{1}{2}\left\{aa^{\dagger},\rho_S\right\}\right] \nonumber \\ 
 &-& d_3(t) \left[a\rho_Sa-\frac{1}{2}\left\{aa,\rho_S\right\}\right] \nonumber \\ 
 &-& d^*_3(t) \left[a^{\dagger}\rho_Sa^{\dagger}-\frac{1}{2}\left\{a^{\dagger}a^{\dagger},\rho_S\right\}\right]
\end{eqnarray}
with five time dependent coefficients $\Omega(t)$, $g(t)$, $d_1(t)$, $d_2(t)$ and $d_3(t)$. The coefficients $\Omega(t)$, $d_1(t)$ and $d_2(t)$ must be real because the master equation must preserve the Hermiticity of the density matrix.
This ansatz yields the following equation for the first moment,
\begin{equation} 
 \frac{d}{dt} \langle a \rangle_t = \left[ -i\Omega(t) + \frac{d_2(t)-d_1(t)}{2} \right] \langle a \rangle_t + g(t),\\
\end{equation}
and the comparison with \eqref{d-moment-a} leads to
\begin{eqnarray}
 \Omega(t) &=&  -\Im \left(\frac{\dot{G}(t)}{G(t)}\right) = \omega_r(t), \label{Omet}\\
 d_2(t) - d_1(t) &=& 2\Re \left(\frac{\dot{G}(t)}{G(t)}\right) = -\gamma(t), \\
 g(t) &=& 0, \label{gt}
\end{eqnarray}
where we have used the definitions \eqref{def-omega_r} and \eqref{def-gamma}.
From the master equation \eqref{ansatz-tcl-meq} we also obtain:
\begin{equation} 
 \frac{d}{dt} \langle a^\dagger a \rangle_t = [d_2(t)-d_1(t)] \langle a^\dagger a \rangle_t + d_2(t).
\end{equation}
The comparison with \eqref{d-moment-a*a} leads to the additional condition
\begin{equation} 
 d_2(t) = \gamma(t) I(t) + \dot{I}(t) = \gamma(t) N(t), \label{d2t}
\end{equation}
where we have used definition \eqref{def-N-t}. This equation also yields
\begin{equation} 
 d_1(t) = d_2(t) + \gamma(t) = \gamma(t)(N(t)+1). \label{d1t}
\end{equation}
Finally, we find from the master equation \eqref{ansatz-tcl-meq}:
\begin{equation} 
 \frac{d}{dt} \langle a a \rangle_t = [-2i\omega_r(t) - \gamma(t)] \langle a a \rangle_t + d_3(t),
\end{equation}
which in view of \eqref{d-moment-aa} gives 
\begin{equation}
 d_3(t)=0. \label{d3t}
\end{equation}
With Eqs.~\eqref{Omet}, \eqref{gt}, \eqref{d2t}, \eqref{d1t} and \eqref{d3t} we 
have determined all five coefficients of the master equation 
\eqref{ansatz-tcl-meq}, and the latter is easily seen to coincide with
the master equation \eqref{tcl-meq}.

\section{Spectral density}\label{app-B}

The spectral density must decrease to zero sufficiently rapidly for $\omega \to 0$ in order for the noise integral $I(t)$ 
to be finite. This can be seen by rewriting $I(t)$, which is defined in Eq.~\eqref{noise-integral}, as follows
\begin{equation}\label{eq:noise_rewritten}
    I(t) = \int_0^\infty d\omega J(\omega) n_E (\omega) \left|  \int_0^t G(t_1) e^{i \omega t_1} d t_1 \right|^2, 
\end{equation}
where $n_E (\omega) = [e^{\omega /k_{\rm B}T} - 1]^{-1}$ is the initial Planck distribution of the 
environmental modes, resulting from the assumption that the bath is initially in a thermal equilibrium
state (see Eq.~\eqref{rho-gibbs}), which behaves as $n_E (\omega)\sim\omega^{-1}$ for small frequencies.
To guarantee that the frequency integral converges for $\omega \to 0$, we
replace the Lorentzian spectral density $J(\omega)$ in \eqref{eq:noise_rewritten} by the spectral density
$\tilde{J}(\omega)$ which shows a linear behavior for low frequencies
$\omega < \omega_m$ below a certain threshold $\omega_m \ll \omega_c,\omega_0$:
\begin{equation}
    \tilde{J}(\omega) = \begin{cases}
    J(\omega_m) \frac{\omega}{\omega_m} &  \text{for} \quad 0 \leq \omega \leq \omega_m \\
    J(\omega) & \text{for} \quad \omega \geq \omega_m
    \end{cases}
\end{equation}
We find that this procedure is consistent as long as the condition
\begin{equation}
    \omega_0 \gg \frac{\gamma_0}{2} \frac{\lambda^2}{\omega_c^2 + \lambda^2}
\end{equation}
holds.
This condition can be obtained by expressing the Laplace transform of $G(t)$ in 
terms of the spectral
density $J(\omega)$ and by requiring that the replacement 
$J(\omega)\to\tilde{J}(\omega)$ has only negligible 
impact on \eqref{eq:noise_rewritten}.


\bibliography{biblio_nonmarkov}

\begin{thebibliography}{54}%
\makeatletter
\providecommand \@ifxundefined [1]{%
 \@ifx{#1\undefined}
}%
\providecommand \@ifnum [1]{%
 \ifnum #1\expandafter \@firstoftwo
 \else \expandafter \@secondoftwo
 \fi
}%
\providecommand \@ifx [1]{%
 \ifx #1\expandafter \@firstoftwo
 \else \expandafter \@secondoftwo
 \fi
}%
\providecommand \natexlab [1]{#1}%
\providecommand \enquote  [1]{``#1''}%
\providecommand \bibnamefont  [1]{#1}%
\providecommand \bibfnamefont [1]{#1}%
\providecommand \citenamefont [1]{#1}%
\providecommand \href@noop [0]{\@secondoftwo}%
\providecommand \href [0]{\begingroup \@sanitize@url \@href}%
\providecommand \@href[1]{\@@startlink{#1}\@@href}%
\providecommand \@@href[1]{\endgroup#1\@@endlink}%
\providecommand \@sanitize@url [0]{\catcode `\\12\catcode `\$12\catcode
  `\&12\catcode `\#12\catcode `\^12\catcode `\_12\catcode `\%12\relax}%
\providecommand \@@startlink[1]{}%
\providecommand \@@endlink[0]{}%
\providecommand \url  [0]{\begingroup\@sanitize@url \@url }%
\providecommand \@url [1]{\endgroup\@href {#1}{\urlprefix }}%
\providecommand \urlprefix  [0]{URL }%
\providecommand \Eprint [0]{\href }%
\providecommand \doibase [0]{http://dx.doi.org/}%
\providecommand \selectlanguage [0]{\@gobble}%
\providecommand \bibinfo  [0]{\@secondoftwo}%
\providecommand \bibfield  [0]{\@secondoftwo}%
\providecommand \translation [1]{[#1]}%
\providecommand \BibitemOpen [0]{}%
\providecommand \bibitemStop [0]{}%
\providecommand \bibitemNoStop [0]{.\EOS\space}%
\providecommand \EOS [0]{\spacefactor3000\relax}%
\providecommand \BibitemShut  [1]{\csname bibitem#1\endcsname}%
\let\auto@bib@innerbib\@empty
\bibitem [{\citenamefont {Abah}\ \emph {et~al.}(2012)\citenamefont {Abah},
  \citenamefont {Ro{\ss}nagel}, \citenamefont {Jacob}, \citenamefont {Deffner},
  \citenamefont {Schmidt-Kaler}, \citenamefont {Singer},\ and\ \citenamefont
  {Lutz}}]{Abah2012}%
  \BibitemOpen
  \bibfield  {author} {\bibinfo {author} {\bibfnamefont {O.}~\bibnamefont
  {Abah}}, \bibinfo {author} {\bibfnamefont {J.}~\bibnamefont {Ro{\ss}nagel}},
  \bibinfo {author} {\bibfnamefont {G.}~\bibnamefont {Jacob}}, \bibinfo
  {author} {\bibfnamefont {S.}~\bibnamefont {Deffner}}, \bibinfo {author}
  {\bibfnamefont {F.}~\bibnamefont {Schmidt-Kaler}}, \bibinfo {author}
  {\bibfnamefont {K.}~\bibnamefont {Singer}}, \ and\ \bibinfo {author}
  {\bibfnamefont {E.}~\bibnamefont {Lutz}},\ }\href {\doibase
  10.1103/PhysRevLett.109.203006} {\bibfield  {journal} {\bibinfo  {journal}
  {Phys. Rev. Lett.}\ }\textbf {\bibinfo {volume} {109}},\ \bibinfo {pages}
  {203006} (\bibinfo {year} {2012})}\BibitemShut {NoStop}%
\bibitem [{\citenamefont {Ro{\ss}nagel}\ \emph {et~al.}(2016)\citenamefont
  {Ro{\ss}nagel}, \citenamefont {Dawkins}, \citenamefont {Tolazzi},
  \citenamefont {Abah}, \citenamefont {Lutz}, \citenamefont {Schmidt-Kaler},\
  and\ \citenamefont {Singer}}]{Rossnagel2016}%
  \BibitemOpen
  \bibfield  {author} {\bibinfo {author} {\bibfnamefont {J.}~\bibnamefont
  {Ro{\ss}nagel}}, \bibinfo {author} {\bibfnamefont {S.~T.}\ \bibnamefont
  {Dawkins}}, \bibinfo {author} {\bibfnamefont {K.~N.}\ \bibnamefont
  {Tolazzi}}, \bibinfo {author} {\bibfnamefont {O.}~\bibnamefont {Abah}},
  \bibinfo {author} {\bibfnamefont {E.}~\bibnamefont {Lutz}}, \bibinfo {author}
  {\bibfnamefont {F.}~\bibnamefont {Schmidt-Kaler}}, \ and\ \bibinfo {author}
  {\bibfnamefont {K.}~\bibnamefont {Singer}},\ }\href {\doibase
  10.1126/science.aad6320} {\bibfield  {journal} {\bibinfo  {journal}
  {Science}\ }\textbf {\bibinfo {volume} {352}},\ \bibinfo {pages} {325}
  (\bibinfo {year} {2016})}\BibitemShut {NoStop}%
\bibitem [{\citenamefont {de~Assis}\ \emph {et~al.}(2019)\citenamefont
  {de~Assis}, \citenamefont {de~Mendon\ifmmode~\mbox{\c{c}}\else \c{c}\fi{}a},
  \citenamefont {Villas-Boas}, \citenamefont {de~Souza}, \citenamefont
  {Sarthour}, \citenamefont {Oliveira},\ and\ \citenamefont
  {de~Almeida}}]{deAssis2019}%
  \BibitemOpen
  \bibfield  {author} {\bibinfo {author} {\bibfnamefont {R.~J.}\ \bibnamefont
  {de~Assis}}, \bibinfo {author} {\bibfnamefont {T.~M.}\ \bibnamefont
  {de~Mendon\ifmmode~\mbox{\c{c}}\else \c{c}\fi{}a}}, \bibinfo {author}
  {\bibfnamefont {C.~J.}\ \bibnamefont {Villas-Boas}}, \bibinfo {author}
  {\bibfnamefont {A.~M.}\ \bibnamefont {de~Souza}}, \bibinfo {author}
  {\bibfnamefont {R.~S.}\ \bibnamefont {Sarthour}}, \bibinfo {author}
  {\bibfnamefont {I.~S.}\ \bibnamefont {Oliveira}}, \ and\ \bibinfo {author}
  {\bibfnamefont {N.~G.}\ \bibnamefont {de~Almeida}},\ }\href {\doibase
  10.1103/PhysRevLett.122.240602} {\bibfield  {journal} {\bibinfo  {journal}
  {Phys. Rev. Lett.}\ }\textbf {\bibinfo {volume} {122}},\ \bibinfo {pages}
  {240602} (\bibinfo {year} {2019})}\BibitemShut {NoStop}%
\bibitem [{\citenamefont {von Lindenfels}\ \emph {et~al.}(2019)\citenamefont
  {von Lindenfels}, \citenamefont {Gr\"ab}, \citenamefont {Schmiegelow},
  \citenamefont {Kaushal}, \citenamefont {Schulz}, \citenamefont {Mitchison},
  \citenamefont {Goold}, \citenamefont {Schmidt-Kaler},\ and\ \citenamefont
  {Poschinger}}]{Lindenfels2019}%
  \BibitemOpen
  \bibfield  {author} {\bibinfo {author} {\bibfnamefont {D.}~\bibnamefont {von
  Lindenfels}}, \bibinfo {author} {\bibfnamefont {O.}~\bibnamefont {Gr\"ab}},
  \bibinfo {author} {\bibfnamefont {C.~T.}\ \bibnamefont {Schmiegelow}},
  \bibinfo {author} {\bibfnamefont {V.}~\bibnamefont {Kaushal}}, \bibinfo
  {author} {\bibfnamefont {J.}~\bibnamefont {Schulz}}, \bibinfo {author}
  {\bibfnamefont {M.~T.}\ \bibnamefont {Mitchison}}, \bibinfo {author}
  {\bibfnamefont {J.}~\bibnamefont {Goold}}, \bibinfo {author} {\bibfnamefont
  {F.}~\bibnamefont {Schmidt-Kaler}}, \ and\ \bibinfo {author} {\bibfnamefont
  {U.~G.}\ \bibnamefont {Poschinger}},\ }\href {\doibase
  10.1103/PhysRevLett.123.080602} {\bibfield  {journal} {\bibinfo  {journal}
  {Phys. Rev. Lett.}\ }\textbf {\bibinfo {volume} {123}},\ \bibinfo {pages}
  {080602} (\bibinfo {year} {2019})}\BibitemShut {NoStop}%
\bibitem [{\citenamefont {Peterson}\ \emph {et~al.}(2019)\citenamefont
  {Peterson}, \citenamefont {Batalh\~ao}, \citenamefont {Herrera},
  \citenamefont {Souza}, \citenamefont {Sarthour}, \citenamefont {Oliveira},\
  and\ \citenamefont {Serra}}]{Peterson2019}%
  \BibitemOpen
  \bibfield  {author} {\bibinfo {author} {\bibfnamefont {J.~P.~S.}\
  \bibnamefont {Peterson}}, \bibinfo {author} {\bibfnamefont {T.~B.}\
  \bibnamefont {Batalh\~ao}}, \bibinfo {author} {\bibfnamefont
  {M.}~\bibnamefont {Herrera}}, \bibinfo {author} {\bibfnamefont {A.~M.}\
  \bibnamefont {Souza}}, \bibinfo {author} {\bibfnamefont {R.~S.}\ \bibnamefont
  {Sarthour}}, \bibinfo {author} {\bibfnamefont {I.~S.}\ \bibnamefont
  {Oliveira}}, \ and\ \bibinfo {author} {\bibfnamefont {R.~M.}\ \bibnamefont
  {Serra}},\ }\href {\doibase 10.1103/PhysRevLett.123.240601} {\bibfield
  {journal} {\bibinfo  {journal} {Phys. Rev. Lett.}\ }\textbf {\bibinfo
  {volume} {123}},\ \bibinfo {pages} {240601} (\bibinfo {year}
  {2019})}\BibitemShut {NoStop}%
\bibitem [{\citenamefont {Myers}\ \emph {et~al.}(2022)\citenamefont {Myers},
  \citenamefont {Abah},\ and\ \citenamefont {Deffner}}]{Myers2022}%
  \BibitemOpen
  \bibfield  {author} {\bibinfo {author} {\bibfnamefont {N.~M.}\ \bibnamefont
  {Myers}}, \bibinfo {author} {\bibfnamefont {O.}~\bibnamefont {Abah}}, \ and\
  \bibinfo {author} {\bibfnamefont {S.}~\bibnamefont {Deffner}},\ }\href
  {\doibase 10.1116/5.0083192} {\bibfield  {journal} {\bibinfo  {journal} {AVS
  Quantum Science}\ }\textbf {\bibinfo {volume} {4}},\ \bibinfo {pages}
  {027101} (\bibinfo {year} {2022})}\BibitemShut {NoStop}%
\bibitem [{\citenamefont {Breuer}\ \emph {et~al.}(2016)\citenamefont {Breuer},
  \citenamefont {Laine}, \citenamefont {Piilo},\ and\ \citenamefont
  {Vacchini}}]{Breuer2016a}%
  \BibitemOpen
  \bibfield  {author} {\bibinfo {author} {\bibfnamefont {H.-P.}\ \bibnamefont
  {Breuer}}, \bibinfo {author} {\bibfnamefont {E.-M.}\ \bibnamefont {Laine}},
  \bibinfo {author} {\bibfnamefont {J.}~\bibnamefont {Piilo}}, \ and\ \bibinfo
  {author} {\bibfnamefont {B.}~\bibnamefont {Vacchini}},\ }\href@noop {}
  {\bibfield  {journal} {\bibinfo  {journal} {Rev. Mod. Phys.}\ }\textbf
  {\bibinfo {volume} {88}},\ \bibinfo {pages} {021002} (\bibinfo {year}
  {2016})}\BibitemShut {NoStop}%
\bibitem [{\citenamefont {Weimer}\ \emph {et~al.}(2008)\citenamefont {Weimer},
  \citenamefont {Henrich}, \citenamefont {Rempp}, \citenamefont {Schr\"oder},\
  and\ \citenamefont {Mahler}}]{Weimer2008}%
  \BibitemOpen
  \bibfield  {author} {\bibinfo {author} {\bibfnamefont {H.}~\bibnamefont
  {Weimer}}, \bibinfo {author} {\bibfnamefont {M.~J.}\ \bibnamefont {Henrich}},
  \bibinfo {author} {\bibfnamefont {F.}~\bibnamefont {Rempp}}, \bibinfo
  {author} {\bibfnamefont {H.}~\bibnamefont {Schr\"oder}}, \ and\ \bibinfo
  {author} {\bibfnamefont {G.}~\bibnamefont {Mahler}},\ }\href {\doibase
  10.1209/0295-5075/83/30008} {\bibfield  {journal} {\bibinfo  {journal} {{EPL}
  (Europhysics Letters)}\ }\textbf {\bibinfo {volume} {83}},\ \bibinfo {pages}
  {30008} (\bibinfo {year} {2008})}\BibitemShut {NoStop}%
\bibitem [{\citenamefont {Esposito}\ \emph {et~al.}(2010)\citenamefont
  {Esposito}, \citenamefont {Lindenberg},\ and\ \citenamefont {den
  Broeck}}]{Esposito2010}%
  \BibitemOpen
  \bibfield  {author} {\bibinfo {author} {\bibfnamefont {M.}~\bibnamefont
  {Esposito}}, \bibinfo {author} {\bibfnamefont {K.}~\bibnamefont
  {Lindenberg}}, \ and\ \bibinfo {author} {\bibfnamefont {C.~V.}\ \bibnamefont
  {den Broeck}},\ }\href {\doibase 10.1088/1367-2630/12/1/013013} {\bibfield
  {journal} {\bibinfo  {journal} {New Journal of Physics}\ }\textbf {\bibinfo
  {volume} {12}},\ \bibinfo {pages} {013013} (\bibinfo {year}
  {2010})}\BibitemShut {NoStop}%
\bibitem [{\citenamefont {Teifel}\ and\ \citenamefont
  {Mahler}(2011)}]{Teifel2011}%
  \BibitemOpen
  \bibfield  {author} {\bibinfo {author} {\bibfnamefont {J.}~\bibnamefont
  {Teifel}}\ and\ \bibinfo {author} {\bibfnamefont {G.}~\bibnamefont
  {Mahler}},\ }\href {\doibase 10.1103/PhysRevE.83.041131} {\bibfield
  {journal} {\bibinfo  {journal} {Phys. Rev. E}\ }\textbf {\bibinfo {volume}
  {83}},\ \bibinfo {pages} {041131} (\bibinfo {year} {2011})}\BibitemShut
  {NoStop}%
\bibitem [{\citenamefont {Alipour}\ \emph {et~al.}(2016)\citenamefont
  {Alipour}, \citenamefont {Benatti}, \citenamefont {Bakhshinezhad},
  \citenamefont {Afsary}, \citenamefont {Marcantoni},\ and\ \citenamefont
  {Rezakhani}}]{Alipour2016}%
  \BibitemOpen
  \bibfield  {author} {\bibinfo {author} {\bibfnamefont {S.}~\bibnamefont
  {Alipour}}, \bibinfo {author} {\bibfnamefont {F.}~\bibnamefont {Benatti}},
  \bibinfo {author} {\bibfnamefont {F.}~\bibnamefont {Bakhshinezhad}}, \bibinfo
  {author} {\bibfnamefont {M.}~\bibnamefont {Afsary}}, \bibinfo {author}
  {\bibfnamefont {S.}~\bibnamefont {Marcantoni}}, \ and\ \bibinfo {author}
  {\bibfnamefont {A.~T.}\ \bibnamefont {Rezakhani}},\ }\href@noop {} {\bibfield
   {journal} {\bibinfo  {journal} {Sci. Rep.}\ }\textbf {\bibinfo {volume}
  {6}},\ \bibinfo {pages} {35568} (\bibinfo {year} {2016})}\BibitemShut
  {NoStop}%
\bibitem [{\citenamefont {Seifert}(2016)}]{Seifert2016}%
  \BibitemOpen
  \bibfield  {author} {\bibinfo {author} {\bibfnamefont {U.}~\bibnamefont
  {Seifert}},\ }\href {\doibase 10.1103/PhysRevLett.116.020601} {\bibfield
  {journal} {\bibinfo  {journal} {Phys. Rev. Lett.}\ }\textbf {\bibinfo
  {volume} {116}},\ \bibinfo {pages} {020601} (\bibinfo {year}
  {2016})}\BibitemShut {NoStop}%
\bibitem [{\citenamefont {Strasberg}\ \emph {et~al.}(2017)\citenamefont
  {Strasberg}, \citenamefont {Schaller}, \citenamefont {Brandes},\ and\
  \citenamefont {Esposito}}]{Strasberg2017}%
  \BibitemOpen
  \bibfield  {author} {\bibinfo {author} {\bibfnamefont {P.}~\bibnamefont
  {Strasberg}}, \bibinfo {author} {\bibfnamefont {G.}~\bibnamefont {Schaller}},
  \bibinfo {author} {\bibfnamefont {T.}~\bibnamefont {Brandes}}, \ and\
  \bibinfo {author} {\bibfnamefont {M.}~\bibnamefont {Esposito}},\ }\href
  {\doibase 10.1103/PhysRevX.7.021003} {\bibfield  {journal} {\bibinfo
  {journal} {Phys. Rev. X}\ }\textbf {\bibinfo {volume} {7}},\ \bibinfo {pages}
  {021003} (\bibinfo {year} {2017})}\BibitemShut {NoStop}%
\bibitem [{\citenamefont {Rivas}(2020)}]{Rivas2020}%
  \BibitemOpen
  \bibfield  {author} {\bibinfo {author} {\bibfnamefont {A.}~\bibnamefont
  {Rivas}},\ }\href {\doibase 10.1103/PhysRevLett.124.160601} {\bibfield
  {journal} {\bibinfo  {journal} {Phys. Rev. Lett.}\ }\textbf {\bibinfo
  {volume} {124}},\ \bibinfo {pages} {160601} (\bibinfo {year}
  {2020})}\BibitemShut {NoStop}%
\bibitem [{\citenamefont {Alipour}\ \emph {et~al.}(2022)\citenamefont
  {Alipour}, \citenamefont {Rezakhani}, \citenamefont {Chenu}, \citenamefont
  {del Campo},\ and\ \citenamefont {Ala-Nissila}}]{Alipour2021}%
  \BibitemOpen
  \bibfield  {author} {\bibinfo {author} {\bibfnamefont {S.}~\bibnamefont
  {Alipour}}, \bibinfo {author} {\bibfnamefont {A.~T.}\ \bibnamefont
  {Rezakhani}}, \bibinfo {author} {\bibfnamefont {A.}~\bibnamefont {Chenu}},
  \bibinfo {author} {\bibfnamefont {A.}~\bibnamefont {del Campo}}, \ and\
  \bibinfo {author} {\bibfnamefont {T.}~\bibnamefont {Ala-Nissila}},\ }\href
  {\doibase 10.1103/PhysRevA.105.L040201} {\bibfield  {journal} {\bibinfo
  {journal} {Phys. Rev. A}\ }\textbf {\bibinfo {volume} {105}},\ \bibinfo
  {pages} {L040201} (\bibinfo {year} {2022})}\BibitemShut {NoStop}%
\bibitem [{\citenamefont {Landi}\ and\ \citenamefont
  {Paternostro}(2021)}]{Landi2021}%
  \BibitemOpen
  \bibfield  {author} {\bibinfo {author} {\bibfnamefont {G.~T.}\ \bibnamefont
  {Landi}}\ and\ \bibinfo {author} {\bibfnamefont {M.}~\bibnamefont
  {Paternostro}},\ }\href {\doibase 10.1103/RevModPhys.93.035008} {\bibfield
  {journal} {\bibinfo  {journal} {Rev. Mod. Phys.}\ }\textbf {\bibinfo {volume}
  {93}},\ \bibinfo {pages} {035008} (\bibinfo {year} {2021})}\BibitemShut
  {NoStop}%
\bibitem [{\citenamefont {Strasberg}\ and\ \citenamefont
  {Esposito}(2019)}]{Strasberg2019}%
  \BibitemOpen
  \bibfield  {author} {\bibinfo {author} {\bibfnamefont {P.}~\bibnamefont
  {Strasberg}}\ and\ \bibinfo {author} {\bibfnamefont {M.}~\bibnamefont
  {Esposito}},\ }\href {\doibase 10.1103/PhysRevE.99.012120} {\bibfield
  {journal} {\bibinfo  {journal} {Phys. Rev. E}\ }\textbf {\bibinfo {volume}
  {99}},\ \bibinfo {pages} {012120} (\bibinfo {year} {2019})}\BibitemShut
  {NoStop}%
\bibitem [{\citenamefont {Colla}\ and\ \citenamefont
  {Breuer}(2021)}]{Colla2021}%
  \BibitemOpen
  \bibfield  {author} {\bibinfo {author} {\bibfnamefont {A.}~\bibnamefont
  {Colla}}\ and\ \bibinfo {author} {\bibfnamefont {H.-P.}\ \bibnamefont
  {Breuer}},\ }\href {\doibase 10.1103/PhysRevA.104.052408} {\bibfield
  {journal} {\bibinfo  {journal} {Phys. Rev. A}\ }\textbf {\bibinfo {volume}
  {104}},\ \bibinfo {pages} {052408} (\bibinfo {year} {2021})}\BibitemShut
  {NoStop}%
\bibitem [{\citenamefont {Zhang}\ \emph {et~al.}(2014)\citenamefont {Zhang},
  \citenamefont {Huang},\ and\ \citenamefont {Yi}}]{Zhang2014}%
  \BibitemOpen
  \bibfield  {author} {\bibinfo {author} {\bibfnamefont {X.~Y.}\ \bibnamefont
  {Zhang}}, \bibinfo {author} {\bibfnamefont {X.~L.}\ \bibnamefont {Huang}}, \
  and\ \bibinfo {author} {\bibfnamefont {X.~X.}\ \bibnamefont {Yi}},\ }\href
  {\doibase 10.1088/1751-8113/47/45/455002} {\bibfield  {journal} {\bibinfo
  {journal} {Journal of Physics A: Mathematical and Theoretical}\ }\textbf
  {\bibinfo {volume} {47}},\ \bibinfo {pages} {455002} (\bibinfo {year}
  {2014})}\BibitemShut {NoStop}%
\bibitem [{\citenamefont {Pozas-Kerstjens}\ \emph {et~al.}(2018)\citenamefont
  {Pozas-Kerstjens}, \citenamefont {Brown},\ and\ \citenamefont
  {Hovhannisyan}}]{Pozas-Kerstjens2018}%
  \BibitemOpen
  \bibfield  {author} {\bibinfo {author} {\bibfnamefont {A.}~\bibnamefont
  {Pozas-Kerstjens}}, \bibinfo {author} {\bibfnamefont {E.~G.}\ \bibnamefont
  {Brown}}, \ and\ \bibinfo {author} {\bibfnamefont {K.~V.}\ \bibnamefont
  {Hovhannisyan}},\ }\href {\doibase 10.1088/1367-2630/aaba02} {\bibfield
  {journal} {\bibinfo  {journal} {New Journal of Physics}\ }\textbf {\bibinfo
  {volume} {20}},\ \bibinfo {pages} {043034} (\bibinfo {year}
  {2018})}\BibitemShut {NoStop}%
\bibitem [{\citenamefont {Thomas}\ \emph {et~al.}(2018)\citenamefont {Thomas},
  \citenamefont {Siddharth}, \citenamefont {Banerjee},\ and\ \citenamefont
  {Ghosh}}]{Thomas2018}%
  \BibitemOpen
  \bibfield  {author} {\bibinfo {author} {\bibfnamefont {G.}~\bibnamefont
  {Thomas}}, \bibinfo {author} {\bibfnamefont {N.}~\bibnamefont {Siddharth}},
  \bibinfo {author} {\bibfnamefont {S.}~\bibnamefont {Banerjee}}, \ and\
  \bibinfo {author} {\bibfnamefont {S.}~\bibnamefont {Ghosh}},\ }\href
  {\doibase 10.1103/PhysRevE.97.062108} {\bibfield  {journal} {\bibinfo
  {journal} {Phys. Rev. E}\ }\textbf {\bibinfo {volume} {97}},\ \bibinfo
  {pages} {062108} (\bibinfo {year} {2018})}\BibitemShut {NoStop}%
\bibitem [{\citenamefont {Pezzutto}\ \emph {et~al.}(2019)\citenamefont
  {Pezzutto}, \citenamefont {Paternostro},\ and\ \citenamefont
  {Omar}}]{Pezzutto2019}%
  \BibitemOpen
  \bibfield  {author} {\bibinfo {author} {\bibfnamefont {M.}~\bibnamefont
  {Pezzutto}}, \bibinfo {author} {\bibfnamefont {M.}~\bibnamefont
  {Paternostro}}, \ and\ \bibinfo {author} {\bibfnamefont {Y.}~\bibnamefont
  {Omar}},\ }\href {\doibase 10.1088/2058-9565/aaf5b4} {\bibfield  {journal}
  {\bibinfo  {journal} {Quantum Science and Technology}\ }\textbf {\bibinfo
  {volume} {4}},\ \bibinfo {pages} {025002} (\bibinfo {year}
  {2019})}\BibitemShut {NoStop}%
\bibitem [{\citenamefont {Mukherjee}\ \emph {et~al.}(2020)\citenamefont
  {Mukherjee}, \citenamefont {Kofman},\ and\ \citenamefont
  {Kurizki}}]{Mukherjee2020}%
  \BibitemOpen
  \bibfield  {author} {\bibinfo {author} {\bibfnamefont {V.}~\bibnamefont
  {Mukherjee}}, \bibinfo {author} {\bibfnamefont {A.~G.}\ \bibnamefont
  {Kofman}}, \ and\ \bibinfo {author} {\bibfnamefont {G.}~\bibnamefont
  {Kurizki}},\ }\href {\doibase 10.1038/s42005-019-0272-z} {\bibfield
  {journal} {\bibinfo  {journal} {Communications Physics}\ }\textbf {\bibinfo
  {volume} {3}},\ \bibinfo {pages} {8} (\bibinfo {year} {2020})}\BibitemShut
  {NoStop}%
\bibitem [{\citenamefont {Wiedmann}\ \emph {et~al.}(2020)\citenamefont
  {Wiedmann}, \citenamefont {Stockburger},\ and\ \citenamefont
  {Ankerhold}}]{Wiedmann2020}%
  \BibitemOpen
  \bibfield  {author} {\bibinfo {author} {\bibfnamefont {M.}~\bibnamefont
  {Wiedmann}}, \bibinfo {author} {\bibfnamefont {J.~T.}\ \bibnamefont
  {Stockburger}}, \ and\ \bibinfo {author} {\bibfnamefont {J.}~\bibnamefont
  {Ankerhold}},\ }\href {\doibase 10.1088/1367-2630/ab725a} {\bibfield
  {journal} {\bibinfo  {journal} {New Journal of Physics}\ }\textbf {\bibinfo
  {volume} {22}},\ \bibinfo {pages} {033007} (\bibinfo {year}
  {2020})}\BibitemShut {NoStop}%
\bibitem [{\citenamefont {Wiedmann}\ \emph {et~al.}(2021)\citenamefont
  {Wiedmann}, \citenamefont {Stockburger},\ and\ \citenamefont
  {Ankerhold}}]{Wiedmann2021}%
  \BibitemOpen
  \bibfield  {author} {\bibinfo {author} {\bibfnamefont {M.}~\bibnamefont
  {Wiedmann}}, \bibinfo {author} {\bibfnamefont {J.~T.}\ \bibnamefont
  {Stockburger}}, \ and\ \bibinfo {author} {\bibfnamefont {J.}~\bibnamefont
  {Ankerhold}},\ }\href {\doibase 10.1140/epjs/s11734-021-00094-0} {\bibfield
  {journal} {\bibinfo  {journal} {The European Physical Journal Special
  Topics}\ }\textbf {\bibinfo {volume} {230}},\ \bibinfo {pages} {851}
  (\bibinfo {year} {2021})}\BibitemShut {NoStop}%
\bibitem [{\citenamefont {Chakraborty}\ \emph {et~al.}(2022)\citenamefont
  {Chakraborty}, \citenamefont {Das},\ and\ \citenamefont {Chru\ifmmode
  \acute{s}\else \'{s}\fi{}ci\ifmmode~\acute{n}\else
  \'{n}\fi{}ski}}]{Chakraborty2022}%
  \BibitemOpen
  \bibfield  {author} {\bibinfo {author} {\bibfnamefont {S.}~\bibnamefont
  {Chakraborty}}, \bibinfo {author} {\bibfnamefont {A.}~\bibnamefont {Das}}, \
  and\ \bibinfo {author} {\bibfnamefont {D.}~\bibnamefont {Chru\ifmmode
  \acute{s}\else \'{s}\fi{}ci\ifmmode~\acute{n}\else \'{n}\fi{}ski}},\ }\href
  {\doibase 10.1103/PhysRevE.106.064133} {\bibfield  {journal} {\bibinfo
  {journal} {Phys. Rev. E}\ }\textbf {\bibinfo {volume} {106}},\ \bibinfo
  {pages} {064133} (\bibinfo {year} {2022})}\BibitemShut {NoStop}%
\bibitem [{\citenamefont {Kaneyasu}\ and\ \citenamefont
  {Hasegawa}(2023)}]{Kaneyasu2023}%
  \BibitemOpen
  \bibfield  {author} {\bibinfo {author} {\bibfnamefont {M.}~\bibnamefont
  {Kaneyasu}}\ and\ \bibinfo {author} {\bibfnamefont {Y.}~\bibnamefont
  {Hasegawa}},\ }\href {\doibase 10.1103/PhysRevE.107.044127} {\bibfield
  {journal} {\bibinfo  {journal} {Phys. Rev. E}\ }\textbf {\bibinfo {volume}
  {107}},\ \bibinfo {pages} {044127} (\bibinfo {year} {2023})}\BibitemShut
  {NoStop}%
\bibitem [{\citenamefont {Ishizaki}\ \emph {et~al.}(2023)\citenamefont
  {Ishizaki}, \citenamefont {Hatano},\ and\ \citenamefont
  {Tajima}}]{Ishizaki2023}%
  \BibitemOpen
  \bibfield  {author} {\bibinfo {author} {\bibfnamefont {M.}~\bibnamefont
  {Ishizaki}}, \bibinfo {author} {\bibfnamefont {N.}~\bibnamefont {Hatano}}, \
  and\ \bibinfo {author} {\bibfnamefont {H.}~\bibnamefont {Tajima}},\ }\href
  {\doibase 10.1103/PhysRevResearch.5.023066} {\bibfield  {journal} {\bibinfo
  {journal} {Phys. Rev. Res.}\ }\textbf {\bibinfo {volume} {5}},\ \bibinfo
  {pages} {023066} (\bibinfo {year} {2023})}\BibitemShut {NoStop}%
\bibitem [{\citenamefont {Liu}\ \emph {et~al.}(2021)\citenamefont {Liu},
  \citenamefont {Jung},\ and\ \citenamefont {Segal}}]{Liu2021}%
  \BibitemOpen
  \bibfield  {author} {\bibinfo {author} {\bibfnamefont {J.}~\bibnamefont
  {Liu}}, \bibinfo {author} {\bibfnamefont {K.~A.}\ \bibnamefont {Jung}}, \
  and\ \bibinfo {author} {\bibfnamefont {D.}~\bibnamefont {Segal}},\ }\href
  {\doibase 10.1103/PhysRevLett.127.200602} {\bibfield  {journal} {\bibinfo
  {journal} {Phys. Rev. Lett.}\ }\textbf {\bibinfo {volume} {127}},\ \bibinfo
  {pages} {200602} (\bibinfo {year} {2021})}\BibitemShut {NoStop}%
\bibitem [{\citenamefont {Colla}\ and\ \citenamefont
  {Breuer}(2022)}]{Colla2022a}%
  \BibitemOpen
  \bibfield  {author} {\bibinfo {author} {\bibfnamefont {A.}~\bibnamefont
  {Colla}}\ and\ \bibinfo {author} {\bibfnamefont {H.-P.}\ \bibnamefont
  {Breuer}},\ }\href {\doibase 10.1103/PhysRevA.105.052216} {\bibfield
  {journal} {\bibinfo  {journal} {Phys. Rev. A}\ }\textbf {\bibinfo {volume}
  {105}},\ \bibinfo {pages} {052216} (\bibinfo {year} {2022})}\BibitemShut
  {NoStop}%
\bibitem [{\citenamefont {Kosloff}\ and\ \citenamefont
  {Rezek}(2017)}]{Kosloff2017}%
  \BibitemOpen
  \bibfield  {author} {\bibinfo {author} {\bibfnamefont {R.}~\bibnamefont
  {Kosloff}}\ and\ \bibinfo {author} {\bibfnamefont {Y.}~\bibnamefont
  {Rezek}},\ }\href {\doibase 10.3390/e19040136} {\bibfield  {journal}
  {\bibinfo  {journal} {Entropy}\ }\textbf {\bibinfo {volume} {19}} (\bibinfo
  {year} {2017}),\ 10.3390/e19040136}\BibitemShut {NoStop}%
\bibitem [{\citenamefont {Fano}(1961)}]{Fano1961}%
  \BibitemOpen
  \bibfield  {author} {\bibinfo {author} {\bibfnamefont {U.}~\bibnamefont
  {Fano}},\ }\href {\doibase 10.1103/PhysRev.124.1866} {\bibfield  {journal}
  {\bibinfo  {journal} {Phys. Rev.}\ }\textbf {\bibinfo {volume} {124}},\
  \bibinfo {pages} {1866} (\bibinfo {year} {1961})}\BibitemShut {NoStop}%
\bibitem [{\citenamefont {Anderson}(1961)}]{Anderson1961}%
  \BibitemOpen
  \bibfield  {author} {\bibinfo {author} {\bibfnamefont {P.~W.}\ \bibnamefont
  {Anderson}},\ }\href {\doibase 10.1103/PhysRev.124.41} {\bibfield  {journal}
  {\bibinfo  {journal} {Phys. Rev.}\ }\textbf {\bibinfo {volume} {124}},\
  \bibinfo {pages} {41} (\bibinfo {year} {1961})}\BibitemShut {NoStop}%
\bibitem [{\citenamefont {Mahan}(2000)}]{Mahan2000}%
  \BibitemOpen
  \bibfield  {author} {\bibinfo {author} {\bibfnamefont {G.~D.}\ \bibnamefont
  {Mahan}},\ }\href@noop {} {\emph {\bibinfo {title} {Many-Particle Physics}}}\
  (\bibinfo  {publisher} {Kluwer Academic/Plenum Publishers},\ \bibinfo
  {address} {New York},\ \bibinfo {year} {2000})\BibitemShut {NoStop}%
\bibitem [{\citenamefont {Huang}\ and\ \citenamefont
  {Zhang}(2022{\natexlab{a}})}]{Huang2022a}%
  \BibitemOpen
  \bibfield  {author} {\bibinfo {author} {\bibfnamefont {W.-M.}\ \bibnamefont
  {Huang}}\ and\ \bibinfo {author} {\bibfnamefont {W.-M.}\ \bibnamefont
  {Zhang}},\ }\href {\doibase 10.1103/PhysRevResearch.4.023141} {\bibfield
  {journal} {\bibinfo  {journal} {Phys. Rev. Res.}\ }\textbf {\bibinfo {volume}
  {4}},\ \bibinfo {pages} {023141} (\bibinfo {year}
  {2022}{\natexlab{a}})}\BibitemShut {NoStop}%
\bibitem [{\citenamefont {Huang}\ and\ \citenamefont
  {Zhang}(2022{\natexlab{b}})}]{Huang2022b}%
  \BibitemOpen
  \bibfield  {author} {\bibinfo {author} {\bibfnamefont {W.-M.}\ \bibnamefont
  {Huang}}\ and\ \bibinfo {author} {\bibfnamefont {W.-M.}\ \bibnamefont
  {Zhang}},\ }\href {\doibase 10.1103/PhysRevA.106.032607} {\bibfield
  {journal} {\bibinfo  {journal} {Phys. Rev. A}\ }\textbf {\bibinfo {volume}
  {106}},\ \bibinfo {pages} {032607} (\bibinfo {year}
  {2022}{\natexlab{b}})}\BibitemShut {NoStop}%
\bibitem [{\citenamefont {Shibata}\ \emph {et~al.}(1977)\citenamefont
  {Shibata}, \citenamefont {Takahashi},\ and\ \citenamefont
  {Hashitsume}}]{Shibata1977}%
  \BibitemOpen
  \bibfield  {author} {\bibinfo {author} {\bibfnamefont {F.}~\bibnamefont
  {Shibata}}, \bibinfo {author} {\bibfnamefont {Y.}~\bibnamefont {Takahashi}},
  \ and\ \bibinfo {author} {\bibfnamefont {N.}~\bibnamefont {Hashitsume}},\
  }\href@noop {} {\bibfield  {journal} {\bibinfo  {journal} {J. Stat. Phys.}\
  }\textbf {\bibinfo {volume} {17}},\ \bibinfo {pages} {171} (\bibinfo {year}
  {1977})}\BibitemShut {NoStop}%
\bibitem [{\citenamefont {Chaturvedi}\ and\ \citenamefont
  {Shibata}(1979)}]{Shibata1979}%
  \BibitemOpen
  \bibfield  {author} {\bibinfo {author} {\bibfnamefont {S.}~\bibnamefont
  {Chaturvedi}}\ and\ \bibinfo {author} {\bibfnamefont {F.}~\bibnamefont
  {Shibata}},\ }\href@noop {} {\bibfield  {journal} {\bibinfo  {journal} {Z.
  Phys. B}\ }\textbf {\bibinfo {volume} {35}},\ \bibinfo {pages} {297}
  (\bibinfo {year} {1979})}\BibitemShut {NoStop}%
\bibitem [{\citenamefont {Hall}\ \emph {et~al.}(2014)\citenamefont {Hall},
  \citenamefont {Cresser}, \citenamefont {Li},\ and\ \citenamefont
  {Andersson}}]{Andersson2014a}%
  \BibitemOpen
  \bibfield  {author} {\bibinfo {author} {\bibfnamefont {M.~J.~W.}\
  \bibnamefont {Hall}}, \bibinfo {author} {\bibfnamefont {J.~D.}\ \bibnamefont
  {Cresser}}, \bibinfo {author} {\bibfnamefont {L.}~\bibnamefont {Li}}, \ and\
  \bibinfo {author} {\bibfnamefont {E.}~\bibnamefont {Andersson}},\ }\href
  {\doibase 10.1103/PhysRevA.89.042120} {\bibfield  {journal} {\bibinfo
  {journal} {Phys. Rev. A}\ }\textbf {\bibinfo {volume} {89}},\ \bibinfo
  {pages} {042120} (\bibinfo {year} {2014})}\BibitemShut {NoStop}%
\bibitem [{\citenamefont {Breuer}(2012)}]{Breuer2012a}%
  \BibitemOpen
  \bibfield  {author} {\bibinfo {author} {\bibfnamefont {H.-P.}\ \bibnamefont
  {Breuer}},\ }\href@noop {} {\bibfield  {journal} {\bibinfo  {journal} {J.
  Phys. B}\ }\textbf {\bibinfo {volume} {45}},\ \bibinfo {pages} {154001}
  (\bibinfo {year} {2012})}\BibitemShut {NoStop}%
\bibitem [{\citenamefont {Breuer}\ and\ \citenamefont
  {Petruccione}(2002)}]{Breuer2002}%
  \BibitemOpen
  \bibfield  {author} {\bibinfo {author} {\bibfnamefont {H.-P.}\ \bibnamefont
  {Breuer}}\ and\ \bibinfo {author} {\bibfnamefont {F.}~\bibnamefont
  {Petruccione}},\ }\href@noop {} {\emph {\bibinfo {title} {The Theory of Open
  Quantum Systems}}}\ (\bibinfo  {publisher} {Oxford University Press},\
  \bibinfo {address} {Oxford},\ \bibinfo {year} {2002})\BibitemShut {NoStop}%
\bibitem [{\citenamefont {van Kampen}(1974{\natexlab{a}})}]{Kampen1974a}%
  \BibitemOpen
  \bibfield  {author} {\bibinfo {author} {\bibfnamefont {N.~G.}\ \bibnamefont
  {van Kampen}},\ }\href@noop {} {\bibfield  {journal} {\bibinfo  {journal}
  {Physica}\ }\textbf {\bibinfo {volume} {74}},\ \bibinfo {pages} {215}
  (\bibinfo {year} {1974}{\natexlab{a}})}\BibitemShut {NoStop}%
\bibitem [{\citenamefont {van Kampen}(1974{\natexlab{b}})}]{Kampen1974b}%
  \BibitemOpen
  \bibfield  {author} {\bibinfo {author} {\bibfnamefont {N.~G.}\ \bibnamefont
  {van Kampen}},\ }\href@noop {} {\bibfield  {journal} {\bibinfo  {journal}
  {Physica}\ }\textbf {\bibinfo {volume} {74}},\ \bibinfo {pages} {239}
  (\bibinfo {year} {1974}{\natexlab{b}})}\BibitemShut {NoStop}%
\bibitem [{\citenamefont {Tu}\ and\ \citenamefont {Zhang}(2008)}]{Tu2008}%
  \BibitemOpen
  \bibfield  {author} {\bibinfo {author} {\bibfnamefont {M.~W.~Y.}\
  \bibnamefont {Tu}}\ and\ \bibinfo {author} {\bibfnamefont {W.-M.}\
  \bibnamefont {Zhang}},\ }\href {\doibase 10.1103/PhysRevB.78.235311}
  {\bibfield  {journal} {\bibinfo  {journal} {Phys. Rev. B}\ }\textbf {\bibinfo
  {volume} {78}},\ \bibinfo {pages} {235311} (\bibinfo {year}
  {2008})}\BibitemShut {NoStop}%
\bibitem [{\citenamefont {Jin}\ \emph {et~al.}(2010)\citenamefont {Jin},
  \citenamefont {Tu}, \citenamefont {Zhang},\ and\ \citenamefont
  {Yan}}]{Jin2010}%
  \BibitemOpen
  \bibfield  {author} {\bibinfo {author} {\bibfnamefont {J.}~\bibnamefont
  {Jin}}, \bibinfo {author} {\bibfnamefont {M.~W.-Y.}\ \bibnamefont {Tu}},
  \bibinfo {author} {\bibfnamefont {W.-M.}\ \bibnamefont {Zhang}}, \ and\
  \bibinfo {author} {\bibfnamefont {Y.}~\bibnamefont {Yan}},\ }\href {\doibase
  10.1088/1367-2630/12/8/083013} {\bibfield  {journal} {\bibinfo  {journal}
  {New Journal of Physics}\ }\textbf {\bibinfo {volume} {12}},\ \bibinfo
  {pages} {083013} (\bibinfo {year} {2010})}\BibitemShut {NoStop}%
\bibitem [{\citenamefont {Lei}\ and\ \citenamefont {Zhang}(2012)}]{Lei2012}%
  \BibitemOpen
  \bibfield  {author} {\bibinfo {author} {\bibfnamefont {C.~U.}\ \bibnamefont
  {Lei}}\ and\ \bibinfo {author} {\bibfnamefont {W.-M.}\ \bibnamefont
  {Zhang}},\ }\href {\doibase https://doi.org/10.1016/j.aop.2012.02.005}
  {\bibfield  {journal} {\bibinfo  {journal} {Annals of Physics}\ }\textbf
  {\bibinfo {volume} {327}},\ \bibinfo {pages} {1408} (\bibinfo {year}
  {2012})}\BibitemShut {NoStop}%
\bibitem [{\citenamefont {Zhang}\ \emph {et~al.}(2012)\citenamefont {Zhang},
  \citenamefont {Lo}, \citenamefont {Xiong}, \citenamefont {Tu},\ and\
  \citenamefont {Nori}}]{Zhang2012}%
  \BibitemOpen
  \bibfield  {author} {\bibinfo {author} {\bibfnamefont {W.-M.}\ \bibnamefont
  {Zhang}}, \bibinfo {author} {\bibfnamefont {P.-Y.}\ \bibnamefont {Lo}},
  \bibinfo {author} {\bibfnamefont {H.-N.}\ \bibnamefont {Xiong}}, \bibinfo
  {author} {\bibfnamefont {M.~W.-Y.}\ \bibnamefont {Tu}}, \ and\ \bibinfo
  {author} {\bibfnamefont {F.}~\bibnamefont {Nori}},\ }\href@noop {} {\bibfield
   {journal} {\bibinfo  {journal} {Phys. Rev. Lett.}\ }\textbf {\bibinfo
  {volume} {109}},\ \bibinfo {pages} {170402} (\bibinfo {year}
  {2012})}\BibitemShut {NoStop}%
\bibitem [{\citenamefont {Sorce}\ and\ \citenamefont
  {Hayden}(2022)}]{Sorce2022}%
  \BibitemOpen
  \bibfield  {author} {\bibinfo {author} {\bibfnamefont {J.}~\bibnamefont
  {Sorce}}\ and\ \bibinfo {author} {\bibfnamefont {P.~M.}\ \bibnamefont
  {Hayden}},\ }\href {\doibase 10.1088/1751-8121/ac65c2} {\bibfield  {journal}
  {\bibinfo  {journal} {J. Phys. A: Math. Theor.}\ } (\bibinfo {year} {2022}),\
  10.1088/1751-8121/ac65c2}\BibitemShut {NoStop}%
\bibitem [{Note1()}]{Note1}%
  \BibitemOpen
  \bibinfo {note} {We remark that an extension of our approach to correlated
  initial states seems possible by means of the formalism developed in \cite
  {Colla2022b}}\BibitemShut {NoStop}%
\bibitem [{\citenamefont {Deffner}\ and\ \citenamefont
  {Lutz}(2008)}]{Deffner2008Feb}%
  \BibitemOpen
  \bibfield  {author} {\bibinfo {author} {\bibfnamefont {S.}~\bibnamefont
  {Deffner}}\ and\ \bibinfo {author} {\bibfnamefont {E.}~\bibnamefont {Lutz}},\
  }\href {\doibase 10.1103/PhysRevE.77.021128} {\bibfield  {journal} {\bibinfo
  {journal} {Phys. Rev. E}\ }\textbf {\bibinfo {volume} {77}},\ \bibinfo
  {pages} {021128} (\bibinfo {year} {2008})}\BibitemShut {NoStop}%
\bibitem [{\citenamefont {Husimi}(1953)}]{Husimi1953Apr}%
  \BibitemOpen
  \bibfield  {author} {\bibinfo {author} {\bibfnamefont {K.}~\bibnamefont
  {Husimi}},\ }\href {\doibase 10.1143/ptp/9.4.381} {\bibfield  {journal}
  {\bibinfo  {journal} {Prog. Theor. Phys.}\ }\textbf {\bibinfo {volume} {9}},\
  \bibinfo {pages} {381} (\bibinfo {year} {1953})}\BibitemShut {NoStop}%
\bibitem [{\citenamefont {Shirai}\ \emph {et~al.}(2021)\citenamefont {Shirai},
  \citenamefont {Hashimoto}, \citenamefont {Tezuka}, \citenamefont {Uchiyama},\
  and\ \citenamefont {Hatano}}]{Shirai2021}%
  \BibitemOpen
  \bibfield  {author} {\bibinfo {author} {\bibfnamefont {Y.}~\bibnamefont
  {Shirai}}, \bibinfo {author} {\bibfnamefont {K.}~\bibnamefont {Hashimoto}},
  \bibinfo {author} {\bibfnamefont {R.}~\bibnamefont {Tezuka}}, \bibinfo
  {author} {\bibfnamefont {C.}~\bibnamefont {Uchiyama}}, \ and\ \bibinfo
  {author} {\bibfnamefont {N.}~\bibnamefont {Hatano}},\ }\href {\doibase
  10.1103/PhysRevResearch.3.023078} {\bibfield  {journal} {\bibinfo  {journal}
  {Phys. Rev. Res.}\ }\textbf {\bibinfo {volume} {3}},\ \bibinfo {pages}
  {023078} (\bibinfo {year} {2021})}\BibitemShut {NoStop}%
\bibitem [{\citenamefont {Latune}\ \emph {et~al.}(2023)\citenamefont {Latune},
  \citenamefont {Pleasance},\ and\ \citenamefont {Petruccione}}]{Latune2023}%
  \BibitemOpen
  \bibfield  {author} {\bibinfo {author} {\bibfnamefont {C.~L.}\ \bibnamefont
  {Latune}}, \bibinfo {author} {\bibfnamefont {G.}~\bibnamefont {Pleasance}}, \
  and\ \bibinfo {author} {\bibfnamefont {F.}~\bibnamefont {Petruccione}},\
  }\href@noop {} {\enquote {\bibinfo {title} {Cyclic quantum engines enhanced
  by strong bath coupling},}\ } (\bibinfo {year} {2023}),\ \Eprint
  {http://arxiv.org/abs/2304.03267} {arXiv:2304.03267 [quant-ph]} \BibitemShut
  {NoStop}%
\bibitem [{\citenamefont {Colla}\ \emph {et~al.}(2022)\citenamefont {Colla},
  \citenamefont {Neubrand},\ and\ \citenamefont {Breuer}}]{Colla2022b}%
  \BibitemOpen
  \bibfield  {author} {\bibinfo {author} {\bibfnamefont {A.}~\bibnamefont
  {Colla}}, \bibinfo {author} {\bibfnamefont {N.}~\bibnamefont {Neubrand}}, \
  and\ \bibinfo {author} {\bibfnamefont {H.-P.}\ \bibnamefont {Breuer}},\
  }\href {\doibase 10.1088/1367-2630/aca709} {\bibfield  {journal} {\bibinfo
  {journal} {New Journal of Physics}\ }\textbf {\bibinfo {volume} {24}},\
  \bibinfo {pages} {123005} (\bibinfo {year} {2022})}\BibitemShut {NoStop}%
\end{thebibliography}%

\end{document}